\documentclass{aa}  
\usepackage[colorlinks=true,     linkcolor=blue, citecolor=blue, filecolor=blue, urlcolor=blue]{hyperref}

\usepackage{graphicx}
\usepackage{float}

\usepackage{txfonts}
\usepackage{placeins}

\newcommand{\zh}{$\rm [Z/H]$}
\newcommand{\mh}{$\rm [M/H]$}
\newcommand{\feh}{$\rm [Fe/H]$}
\newcommand{\cfe}{$\rm [C/Fe]$}
\newcommand{\omg}{$\rm [O/Mg]$}
\newcommand{\mgfe}{$\rm [Mg/Fe]$}
\newcommand{\ofe}{$\rm [O/Fe]$}
\newcommand{\xfe}{$\rm [X/Fe]$}
\newcommand{\xh}{$\rm [X/H]$}
\newcommand{\nafe}{$\rm [Na/Fe]$}
\newcommand{\nfe}{$\rm [N/Fe]$}
\newcommand{\cafe}{$\rm [Ca/Fe]$}
\newcommand{\tife}{$\rm [Ti/Fe]$}
\newcommand{\sife}{$\rm [Si/Fe]$}
\newcommand{\tioi}{$\rm TiO1$}
\newcommand{\atio}{$\rm aTiO$}
\newcommand{\tioiir}{$\rm TiO2_{SDSS}$}
\newcommand{\tioii}{$\rm TiO2$}

\newcommand{\cafr}{$\rm Ca4227r$}
\newcommand{\cai}{$\rm Ca1$}
\newcommand{\caii}{$\rm Ca2$}
\newcommand{\caiii}{$\rm Ca3$}

\newcommand{\hbo}{$\rm H\beta_o$}
\newcommand{\hb}{$\rm H\beta$}

\newcommand{\hgf}{$\rm H\gamma_F$}

\newcommand{\hdf}{$\rm H\delta_F$}
\newcommand{\mgb}{$\rm Mgb5177$}
\newcommand{\nai}{$\rm NaD$}
\newcommand{\naii}{$\rm NaI8190$}

\newcommand{\cnii}{$\rm CN2$}

\newcommand{\mgi}{$\rm Mg1$}

\newcommand{\afe}{$\rm [\alpha/Fe]$}
\newcommand{\kms}{\,km\,s$^{-1}$}

\newcommand{\gammab}{$\rm \Gamma_b$}

\newcommand{\mgf}{$\rm Mg4780$}

\begin{document}
\nolinenumbers
\titlerunning{Abundance ratios for the M31 bulge}
\authorrunning{F. La Barbera et al.}

   \title{Elemental abundance ratios for the bulge of M31}


   \author{F. La Barbera~\inst{1}, 
   A. Vazdekis~\inst{2, 3}, F. Matteucci~\inst{4, 5, 6}, E. Spitoni~\inst{5}, 
A. Pasquali~\inst{7}, I. Mart\'in-Navarro~\inst{2, 3} 
          }

   \institute{INAF-Osservatorio Astronomico di Capodimonte, sal. Moiariello
16, Napoli, 80131, Italy\\
              \email{francesco.labarbera@inaf.it}
         \and
   Instituto de Astrof\'\i sica de Canarias, Calle V\'\i a L\'actea s/n, E-38205
         \and
   Departamento de Astrof\'\i sica, Universidad de La Laguna (ULL), E-38206  La Laguna, Tenerife, Spain
         \and
    Dipartimento di Fisica, Sezione di Astronomia, Universit{\`a} di Trieste, Via G. B. Tiepolo 11, I-34143 Trieste, Italy
          \and
    INAF-Osservatorio Astronomico di Trieste, via G.B. Tiepolo 11, 34131, Trieste, Italy
          \and
    INFN-Sezione di Trieste, via Valerio 2, 34134 Trieste, Italy
        \and
   Astronomisches Rechen-Institut, Zentrum f\"ur Astronomie, Universit\"at Heidelberg, M\"onchhofstr. 12-14, D-69120 Heidelberg, Germany
   }

   \date{Received ; accepted}

   \abstract{We present radial trends of metallicity (\feh) and abundance ratios ($\rm [X/Fe]$) for several chemical elements -- including  C, N, Na, and the so-called $\alpha$-elements (O, Mg, Si, Ca, and Ti) -- in the bulge of M31, out to a projected galactocentric distance of $\sim 0.6$~kpc. We estimated abundances using multiple approaches,  including full-spectrum fitting, full-index fitting, and line-strength analysis, in combination with different stellar population models. We first tested these techniques on mock spectra and SDSS stacked spectra of early-type galaxies (ETGs), and then applied them to high-quality long-slit spectroscopy of the M31 bulge obtained with the OSIRIS spectrograph at the { Gran Telescopio CANARIAS}.    
   We find that O, N, and Na are significantly enhanced relative to Fe across the bulge, with typical abundances $\gtrsim $0.3~dex. In particular, N and Na show steep central enhancements, reaching $\sim 0.5$~dex. C, Mg, and Si exhibit intermediate enhancements of $\rm [X/Fe] \sim 0.2$~dex, with C and Mg decreasing toward the center to $\lesssim $0.1~dex; while Ca, and to a lesser extent Ti, closely follow Fe, with $\rm [X/Fe] < 0.1$~dex within uncertainties. Applying the same analysis  to SDSS stacked spectra of ETGs as a function of velocity dispersion revealed that the abundance pattern of the M31 bulge closely resembles that of the most massive galaxies, except for N, which is significantly more enhanced (by $\sim 0.1$~dex) in the bulge. 
   For the bulk of the bulge, chemical evolution models assuming high star-formation efficiency and a short gas infall timescale reproduce the overall trends in \feh\ and \xfe . In the central region ($\lesssim 100$~pc), the high metallicity content of the bulge can be explained by either an Initial Mass Function flatter than Salpeter at high mass, or a prolonged star formation. Additional processes,  such as differential galactic winds, appear necessary to account for the observed decoupling among $\alpha$ elements and the strong central N enhancement. 
   Our results support a scenario whereby the bulk of the M31 bulge formed during a fast and intense episode of star formation.
   }


   \maketitle

\nolinenumbers

\section{Introduction}
\label{sec:intro}

Understanding the stellar population properties of galaxies—such as age, metallicity, the initial mass function (IMF), and chemical abundance ratios—is fundamental for constraining models of galaxy formation and evolution. In particular, abundance ratios serve as key diagnostics of star-formation timescales and nucleosynthetic processes, as both are sensitive to the mass distribution of stars (due to the distinct progenitors of various elements) and to stellar lifetimes.

In systems with unresolved stellar populations, such as most galaxies beyond the Local Group, abundance ratios must be inferred through detailed spectral analysis. This requires high signal-to-noise (S/N) integrated-light spectra and comparison with stellar population synthesis models that account for variations in individual elemental abundances~\citep{TMB:03, CvD12a, Vazdekis:15, LB:17, Knowles:2023}. 
Stellar population parameters can be derived using various methods, including full-spectrum fitting ({e.g., \citealt{RW:92, CE:04, CID:05, Chilingarian2007, Benitez2012, CvD12b, Asad2013}}); fitting restricted to selected spectral regions containing key absorption features, hereafter referred to as full-index fitting (e.g., \citealt{NMN:19}); and the analysis of line strengths for well-defined sets of spectral indices, such as the Lick { system~\citep{Worthey1997, Trager:2000, Leonardi2003, Thomas:2005, Serven:2005, Schiavon:07, Graves2008}}.
Because the effects of abundance variations on spectral features are subtle -- typically at the few percent level -- it is crucial to validate these techniques against nearby systems, for which chemical abundances can be independently constrained via resolved stellar population studies ({see, e.g., \citealt{Chilingarian2018, Asad2022}}). With the advent of high-resolution facilities such as the James Web Space Telescope (JWST) and, in the near future, the Extremely Large Telescope (ELT), nearby galaxies, such as M31,  will offer unique opportunities in this regard.
  
Owing to its proximity, the bulge of the Andromeda galaxy (M31) allows for the acquisition of high S/N spectra at a relatively low observational cost. Moreover, its stellar populations share key characteristics with those of massive early-type galaxies (ETGs): they are old, alpha enhanced, and exhibit super-solar metallicity {in the central region} (e.g., \citealt{Saglia:2010}). As such, M31 provides a natural laboratory for applying and testing stellar population analysis methods commonly used in the study of ETGs.

The nature of stellar populations in the bulge of M31 has been debated since the 1970s. Early studies led to conflicting interpretations: \citet{SpinTa:71} interpreted strong Na I absorption near 8200 \AA\ as evidence for a dwarf-rich (i.e., bottom-heavy) IMF, while \citet{Faber:1972} argued, based on CO absorption at 2.3 $\mu$m, for a population dominated by giant stars. More recently, \citet{CvD12b} 
analyzed spectra  of the central few arc seconds of M31 and found evidence of an IMF normalization intermediate between those of Kroupa and Salpeter. \citet[hereafter LB21]{LB:21} performed the first radially resolved study of the IMF in the M31 bulge, showing that its innermost $\sim$10'' region exhibits a mildly bottom-heavy IMF, while for most of the bulge, the low-mass end of the IMF is consistent with a Kroupa-like distribution, as expected from the relatively modest velocity dispersion of the bulge ($\sigma \sim 150$~\kms ; {e.g.,~\citealt{Saglia:2010}}). 

Surprisingly, despite extensive studies of its global stellar population properties, the detailed chemical abundance patterns in the M31 bulge remain poorly constrained. Notably, \citet{CvD12b} reported a central sodium abundance as high 
as \nafe $\sim 1$~dex -- significantly exceeding the values observed in the Milky Way (MW) bulge for dwarf stars~\citep{Bensby:17}, though possibly closer to those measured in more metal-rich giant stars~\citep{Lec:2007}. Furthermore, \citet{Z:15} showed that different stellar population models yield diverging predictions for the Na abundance in the bulge,  emphasizing  the need for improved modeling and observations.
The $\alpha$-element abundance, \afe , in the M31 bulge was derived { by~\citet{Saglia:2010} and LB21}, who both found an enhancement of $\sim$0.2~dex, with a decrease in the innermost few arc seconds. It is worth noting that both studies assumed all $\alpha$-elements to vary in lockstep, but based their estimates primarily on Mg and Fe absorption features. Therefore, their results likely reflect the radial behavior of \mgfe. Indeed, individual elements may exhibit different radial trends. This is supported by observations of spheroidal systems such as ETGs, where not all elements closely follow each other (e.g.,~\citealt{JTM:2012, CvD12b}). 

Chemical evolution models provide a powerful tool for interpreting the abundance patterns and metal content of both resolved and unresolved stellar populations. 
Based on assumptions about the star-formation rate (SFR) and efficiency, gas flows, and the IMF, these models predict the time evolution of various chemical properties, which can be directly compared with observations. The assumed stellar yields play a central role in producing these predictions. 
{ \citet{MU:2015}}  computed a set of chemical evolution models for the bulge of M31, finding that the model that best reproduces the metallicity distribution function (MDF) of bulge dwarf stars { from~\citet{SJ:2005}} assumes a Salpeter IMF and a short, intense burst of star formation.
Although { \citet{MU:2015}} provided predictions for several abundances in the bulge, no detailed comparison with observations has been possible so far, due to the lack of precise abundance determinations for the M31 bulge. 

In this work, we present a detailed determination of abundance ratios
for individual chemical elements in the M31 bulge as a function of
galactocentric distance, out to $\sim 0.6$~kpc from the center. For
this purpose, we used high-quality long-slit spectroscopy along the
bulge’s major axis, obtained with the Optical System for
Imaging and low-Intermediate-Resolution Integrated Spectroscopy (OSIRIS)
instrument at the Gran Telescopio CANARIAS (GTC).  These data were
first presented in LB21, where they were used to constrain the
low-mass end of the IMF across the bulge.  A further goal of the
present study is to validate and compare different techniques for
deriving elemental abundances, making use of both mock spectra and
stacked spectra of ETGs from the Sloan Digital Sky Survey (SDSS). The
comparison between the M31 bulge and ETGs also provides key insights
into the bulge’s formation history. To this end, we performed a detailed
analysis of the radial profiles of metallicity and abundance ratios,
confronting them with predictions from an updated version of the
chemical evolution models presented { in~\citet{MU:2015}}.

The structure of the paper is as follows.  In Sec.~\ref{sec:data}, we
describe the data used for the stellar population analysis of the M31
bulge (Sec.~\ref{sec:data_m31}) and of ETGs as a function of velocity
dispersion (Sec.~\ref{sec:data_etgs}).  The stellar population models
adopted in the analysis are presented in Sec.~\ref{sec:models}, while
in Sec.~\ref{sec:methods} we outline the methods used to estimate elemental
abundance ratios. In Sec.~\ref{sec:cmodels}, we describe the chemical
evolution models for the M31 bulge.  Results are presented in
Sec.~\ref{sec:results} and discussed in
Sec.~\ref{sec:discussion}. Throughout this work, we assume a distance
of 785~kpc to M31~\citep{McConnachie:2005}, corresponding to a
physical scale of $\rm \sim 3.8$ pc/arcsec.

\section{Data}
\label{sec:data}
\subsection{Optical spectroscopy for the bulge of M31}
\label{sec:data_m31}
{ Long-slit spectroscopy along the major axis of M31’s bulge was obtained with OSIRIS at the GTC (Nasmyth-B focus, Roque de los Muchachos Observatory). The slit was centered on the bulge at RA=00:42:44.57 and Dec=+41:16:05.7, with a position angle of 48~deg (see LB21). We summarize here the main data properties and refer to LB21 for further details. Observations employed the R2500 grisms in the $\rm U$, $\rm V$, $\rm R$, and $\rm I$ bands, using a 0.4\arcsec\ slit that provided a spectral resolution of $\sigma \sim 38$~\kms\ and a spatial scale of $0.254$\arcsec\,pixel$^{-1}$, reaching galactocentric distances up to $\sim 200$\arcsec\ ($\sim 760$~pc). The typical seeing was $\sim 1$\arcsec Full Width at Half Maximum (FWHM).}  We corrected the spectra along the slit for radial velocity, and radially binned { them} along both sides of the slit, as detailed in LB21. In short, the central bin was set to be 1.5\arcsec\ wide around the photometric center of the galaxy, while  for the other bins we adapted the bin width to ensure a minimum signal-to-noise ratio ($\rm S/N$) of 70 per \AA . This procedure resulted in a set of 51 binned spectra, spanning the radial range from approximately $-200$\arcsec\ to $+150$\arcsec\ around the photometric center of the bulge.~\footnote{Since the center of the bulge was slightly offset along the slit, the "negative" side of the slit reaches a larger galactocentric distance than the "positive" side.}  Examples of the binned spectra are shown in figure~1 {of LB21}. 
The data were also corrected for interstellar medium (ISM) contamination (from both the Milky Way and M31) in the NaD absorption at $\lambda \sim 5900$\AA\ (see Appendix~B of LB21). The bulge velocity dispersion is $\sim 150$–160~\kms, rising to $\sim 200$~\kms\ in the innermost few arc seconds (see figure~E1 of LB21). In the present analysis, each binned spectrum, characterized by a given $\sigma$, is compared to stellar population model predictions (see below) that have been smoothed to the same $\sigma$.

\subsection{SDSS stacked spectra}
\label{sec:data_etgs}
We relied on a set of 18 stacked spectra of ETGs with velocity dispersion, $\sigma$, in the range from $\sim 100$ to $\sim 300$~\kms . The ETGs were drawn from the SPIDER sample (\citealt{SpiderI}), consisting of $\sim 40,000$ nearby ($0.05 < z < 0.095$) ETGs with spectra available from Data Release 6 of the SDSS~\citep{SDSS:DR6}. SPIDER ETGs are defined as bulge-dominated systems with passive spectra within the SDSS fibers. The spectra cover the wavelength range 3800--9200$\AA$, with a spectral resolution of $\sim$70~\kms.

The SDSS stacks were constructed as detailed { in~\citet{LB:13}} and~\citet{F:13},  using a subsample of 24,781 SPIDER ETGs, with 
$100 \le \sigma \le 320$~\kms , low internal extinction (i.e., a color excess $\rm E(B-V) < 0.1$), and S/N ratios above the first quartile of the distribution in each $\sigma$ bin.
The stacking was performed in 18 narrow bins of $\sigma$, each with a width of 10\kms, except for the two highest bins, defined as $[260, 280]$ and $[280, 320]$\kms, respectively. Most stacked spectra reach an S/N of several hundred across the full SDSS wavelength range (see table~1 { of \citet{LB:13}}).

\section{Stellar population models}
\label{sec:models}
{ Our analysis is based on E-MILES~\citep[for direct comparison with LB21]{Vazdekis:2016} and \citet[hereafter CvD18; for their theoretical abundance-ratio responses, see below]{CvD18} stellar population models}.~\footnote{{ Both rely on the MILES stellar library, which is one of  the most widely used optical library in stellar population synthesis.}}

The E-MILES  simple stellar population (SSP) models\footnote{{In the optical, the E-MILES models coincide with the ``MILES'' models of~\citet{Vazdekis:10} and, for the metallicity range of interest ($\rm [Z/H] \gtrsim 0$), closely match scaled-solar SSP models, such as~\citet{Vazdekis:15} and~\citet[sMILES]{Knowles:2023}.} } span the spectral range from $0.168$ to 5~$\mu$m, and are based on the NGSL~\citep{Gregg:2006}, MILES~\citep{MILESI}, Indo-US~\citep{Valdes04}, CaT~\citep{CATI} and IRTF~\citep{IRTFI,IRTFII} empirical stellar libraries (see also~\citealt{Vazdekis:12},~\citealt{RV:16}). 
We use E-MILES models computed for two sets of scaled-solar theoretical isochrones: those of \citet{Padova00} (Padova00; hereafter ``iP'') and \citet{Pietrinferni04} (BaSTI; hereafter ``iT''), the latter having lower temperatures  at the low-mass end~(see \citealt{Vazdekis:15}, and references therein). The SSPs are computed for ages from $\sim 0.06$ to $\sim 17.8$\,Gyr ($0.03$ to $14$~Gyr), and metallicity, \zh, from $-2.2$ to $+0.22$~dex ($-2.27$ to $+0.26$~dex), for Padova00 (BaSTI)~\footnote{BaSTI models are also available for \zh$=0.4$ (see \citealt{Vazdekis:2016}), but these models have lower quality, and an effective metallicity of $0.32$~dex for old SSPs, and hence are not used in the present work.} isochrones. We use models computed for a ``bimodal'' initial mass function (IMF), i.e., a single power-law distribution whose logarithmic slope, $\rm \Gamma_b$, is tapered at the low-mass end ($\lesssim 0.6 \, M_\odot$). In this parametrization, increasing the high-mass end slope does also increase the dwarf-to-giant ratio in the IMF (implying a more bottom-heavy distribution) through its overall normalization. For \gammab$=1.3$, the bimodal IMF is very similar to the Kroupa IMF. 

The CvD18 models are an updated version of those by~\citet{CvD12a}. They cover the spectral range from 0.35 to 2.5~$\mu$m and are based on the MILES and extended IRTF~\citep{Villaume:2017} empirical stellar libraries, in the optical and NIR, respectively. The models adopt MIST isochrones~\citep{Choi:2016, Dotter:2016}, and span ages from 1 to 13.5~Gyr, and metallicities from -1.5 to 0.2~dex. The SSPs adopt a three-segment IMF, with variable slopes, $\rm x_1$ between 0.1--0.5~$\rm M_\odot$, $\rm x_2$ between 0.5--1~$\rm M_\odot$, and a fixed Salpeter slope at higher masses.

CvD18 also provide a set of theoretical SSPs, computed for a Kroupa IMF, at ages of $1, 3, 5, 9$ and 13~Gyr, and metallicities of $-1.5$,  $-1$, $-0.5$, $0$, and $+0.2$~dex, where the abundance of individual elements is varied independently. In this work, we focus on the elements that produce the most significant variations in the optical spectral range, namely the $\alpha$-elements (O, Mg, Si, Ca, Ti), as well as C, N, and Na. { We note} that CvD18 models vary O, Ne, and S in lockstep, denoting the resulting abundance ratio as [$\alpha$s/Fe]. Here, we { adopt} the same convention, assuming that O dominates the spectral response, i.e., [O/Fe]=[$\alpha$s/Fe].
For each element X, CvD18 compute theoretical SSP spectra, $\rm S_X$, for reference abundance ratios $\rm [X/H]_r=\pm0.3$~dex, with the exceptions of C (varied by $\rm \pm 0.15$~dex), and Na (varied by $\pm 0.3, +0.6$ and $+0.9$~dex, respectively). We estimated the spectral response to a given abundance ratio \xh\ through linear interpolation:
\begin{equation}
\rm R_{X} = 1 + \frac{[X/H]}{[X/H]_r} \cdot \left( \frac{S_X}{S_{\odot}} - 1 \right) ,
\label{eq:resp}
\end{equation}
where $\rm S_{\odot}$ is the scaled-solar theoretical spectrum. For a given age and metallicity, $\rm R_X$ is obtained by linear interpolation over the grid of CvD18 theoretical models (see above). 

Multiplying the responses $\rm R_X$ with the empirical SSP spectrum (either E-MILES or CvD18) allowed us to correct it to a given abundance pattern. We note that stellar libraries underlying both the E-MILES and CvD18 SSPs follow the abundance pattern of the Milky Way (MW), i.e., they are approximately scaled-solar at solar and super-solar metallicity, while becoming significantly $\alpha$--enhanced below $\sim -0.2$~dex.
In practice,  all spectra analyzed in this work (i.e., those of the M31 bulge and the stacked ETG spectra) have metallicities above this threshold. Therefore, no correction for the MW pattern is required when inferring \xfe\ abundances.  The only exception is oxygen, as in the solar neighborhood \ofe\ decreases with metallicity, becoming negative in the super-solar metallicity regime~\citep{Bensby:2004}, which is relevant for the innermost region of M31 and for most massive ETGs. We further discuss this issue in Sec.~\ref{sec:results}.

\begin{figure}
\begin{center}
 \leavevmode
 \includegraphics[width=8.cm]{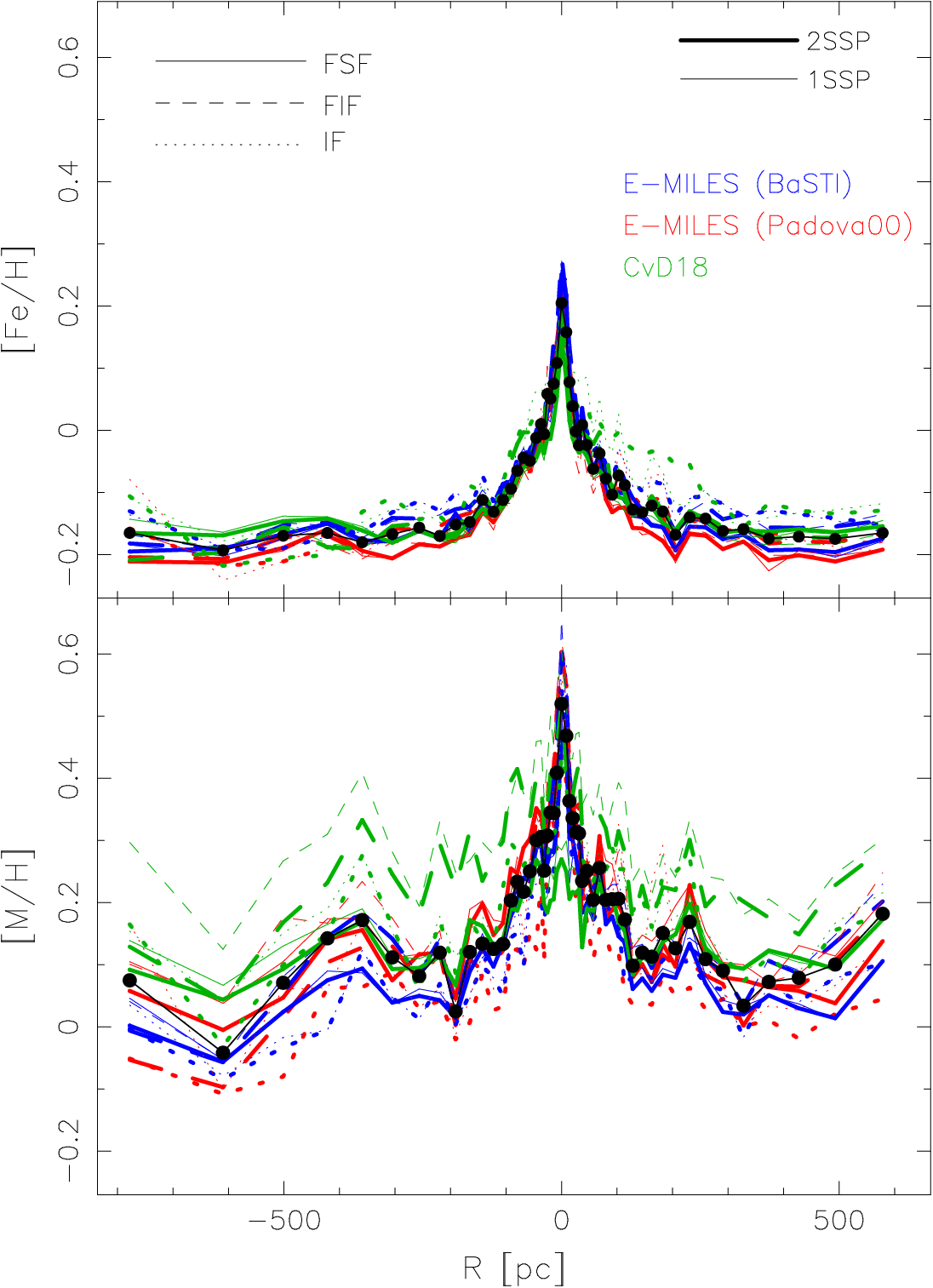}
\end{center}
 \caption{
Metallicity as a function of galactocentric distance, R, for the bulge of M31. Positive and negative values of R correspond to opposite sides of the slit. The top and bottom panels show \feh\ and \mh , respectively.  
 Lines with different colors refer to E-MILES BaSTI (blue) and Padova00 (red), and Cvd18 models (green), respectively. Thick and thin lines indicate  1SSP and 2SSP models, respectively, while solid, dashed, and dotted lines plot results of different methods, i.e., full-spectral fitting, full-index fitting, and index fitting, respectively, as shown by the labels in the top panel. Median trends are plotted as black curves with black dots.
 }
   \label{fig:xh_m31}
\end{figure}

\begin{figure}
\begin{center}
 \leavevmode
 \includegraphics[width=8.cm]{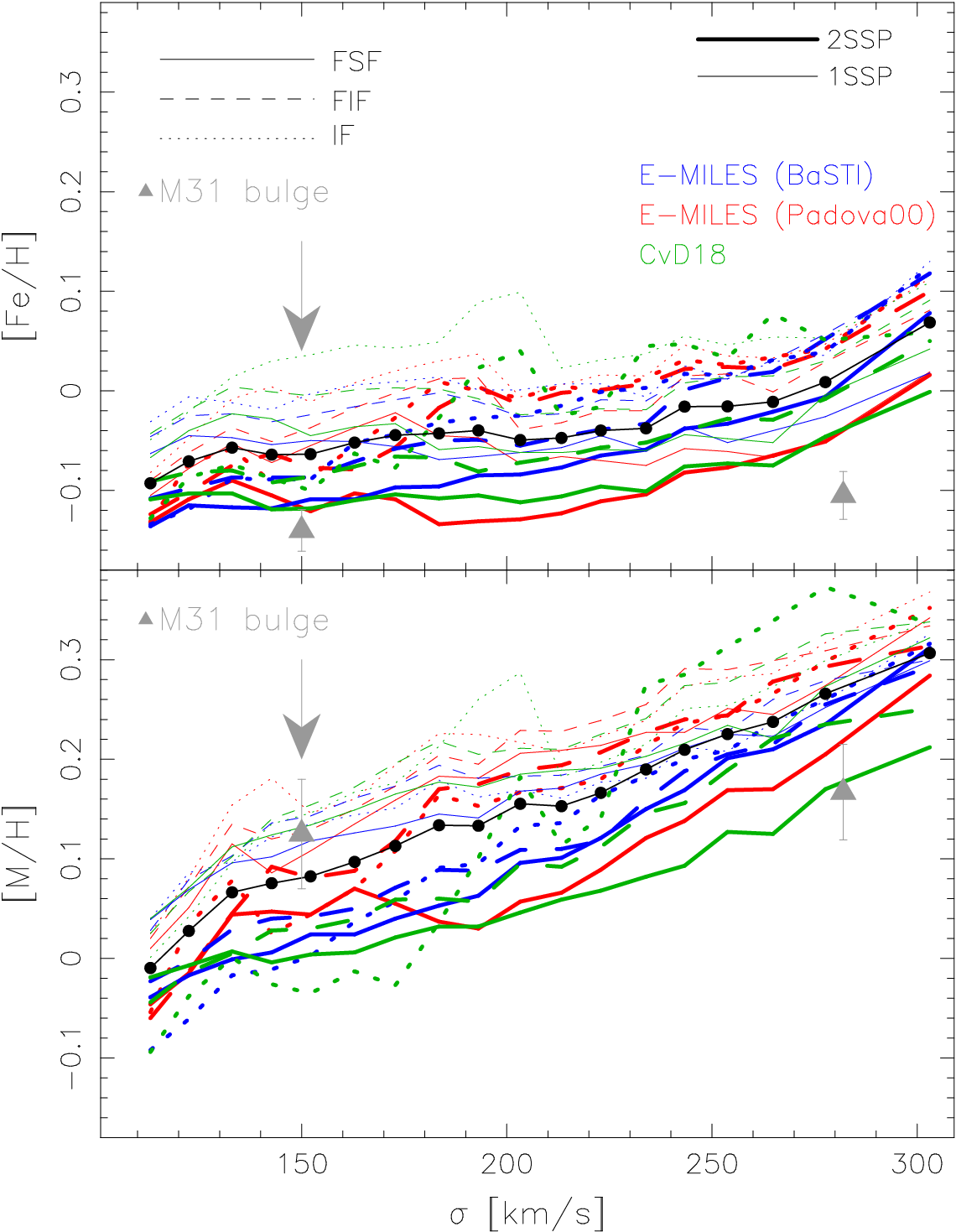}
\end{center}
 \caption{
 Same as Fig.~\ref{fig:xh_m31}, but plotting the metallicity profiles of ETGs stacked spectra as a function of velocity dispersion $\sigma$. 
The velocity dispersion of the M31 bulge ($\sigma \sim 150$~\kms ) is marked with a vertical gray arrow in both panels. The gray triangles with error bars, at $\sigma \sim 150$~\kms\ and $\sigma \sim 280$~\kms\, show the estimates of \feh\ and \mh\ obtained by mimicking the Sloan fiber aperture for low- and high-$\sigma$ ETGs (see the text). 
 }
   \label{fig:xh_sdss}
\end{figure}

\begin{figure*}
\begin{center}
 \leavevmode
 \includegraphics[width=18cm]{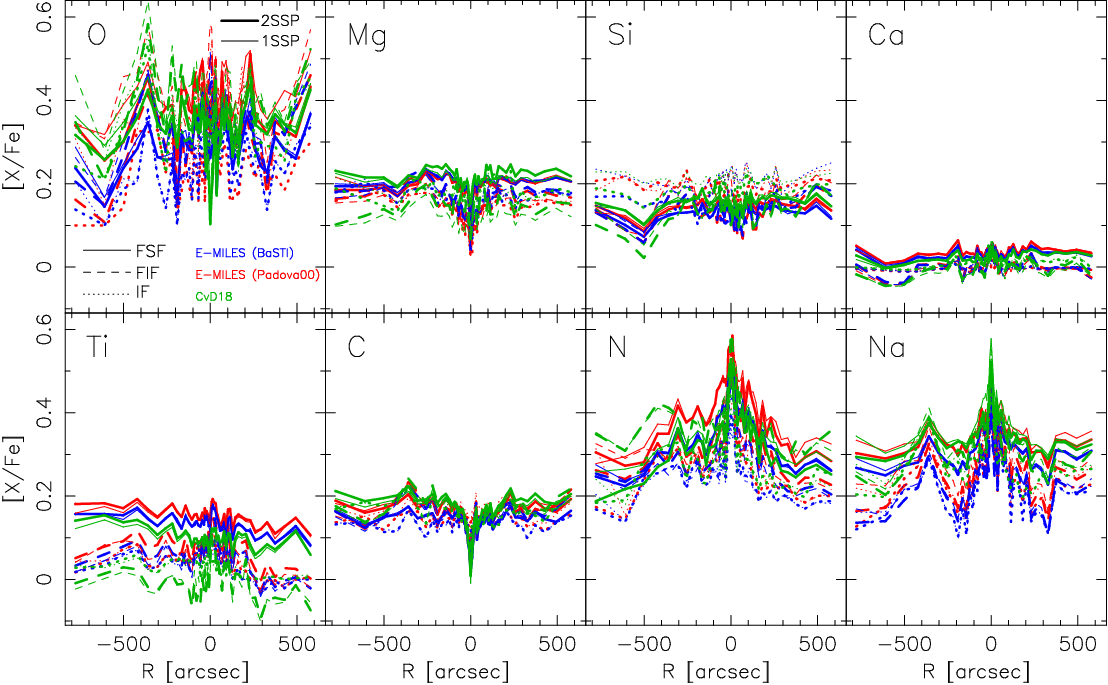}
\end{center}
 \caption{
 Individual abundance ratios as a function of galactocentric distance, R, for the bulge of M31. Positive and negative values of R refer to different sides of the slit. From left to right, and top to bottom, { $\alpha$ elements} elements are shown first, namely O, Mg, Si, Ca, and Ti (in order of decreasing atomic number), followed by C, N, and Na. 
 Lines with different colors refer to E-MILES BaSTI (blue) and Padova00 (red), and Cvd18 models (green), respectively. Thick and thin lines are for 1SSP and 2SSP models, respectively, while solid, dashed, and dotted lines plot results of different methods, i.e., full-spectral fitting, full-index fitting, and index fitting, respectively, as shown by the labels in the top--left panel.
 }
   \label{fig:xfe_m31}
\end{figure*}

\section{Methods to estimate abundance ratios}
\label{sec:methods}
Abundance ratios are commonly derived either by directly fitting galaxy spectra in wavelength space (e.g.,~\citealt{V:99, CvD12b}), or by fitting the strengths of specific (Lick-like) absorption features (e.g.,~\citealt{Worthey:1994}; { \citealt{JTM:2012}}). Each approach has its own advantages and limitations, and is affected by different sources of uncertainty stemming from the adopted stellar population models. To mitigate these effects, we adopt different techniques to estimate the \xfe\ abundance ratios, combining results into final estimates.
Specifically, we measure elemental abundance ratios using three different methods, namely full spectral fitting (FSF), full-index fitting (FIF), and index fitting (IF). 

For the FSF, given the observed spectrum, ${\rm} O_{\lambda}$, with its associated uncertainty, $\sigma_{\lambda}$ (see Sec.~\ref{sec:data}), we determined the best-fitting stellar population model, $M_{\lambda}$, by minimizing the following expression:
\begin{equation}
\chi^2 = \sum_\lambda \frac{ \left( O_{\lambda} - M_{\lambda} \right)^2 }{\sigma_{\lambda}^2}.
\label{eq:sfit}
\end{equation}
The model spectrum, $\rm M_{\lambda}$, is defined as the sum of one or
two SSPs { (see below)}, each with given age, $\rm Age_i$,
metallicity, $\rm [Z/H]_i$, and mass fraction, $\rm f_i$.  Elemental
abundance variations, \xfe, with $X = { {\rm O, Mg, Si, Ca, Ti, C, N,
    Na} }$, are incorporated by applying response functions to each
SSP spectrum (see Eq.~\ref{eq:resp}).  Following the same approach as
in~\citet{CvD12b}, the fit is performed by splitting the spectra into
four wavelength intervals~\footnote{{ The lower limit ($\sim 4000$~\AA )
reflects increased uncertainty in abundance-ratio responses at shorter
wavelengths, while the upper limit ($8750$~\AA ) matches the SDSS
stacked-spectra coverage (Sec.~\ref{sec:data_etgs}).  The same results
are obtained when fitting a single wavelength interval, but using four
intervals permits a lower-degree polynomial, which improves the fit
while accounting for continuum shape mismatches.}  }:
$\Delta(\lambda)=$4000--4800~\AA, 4800--5800~\AA, 5800--6400~\AA, and
8000--8750~\AA .  For each interval, to maximize the information
extracted from absorption features and account for continuum shape
mismatches, we include a multiplicative polynomial in the fit, with
degree $\Delta(\lambda)/(100$\AA$)$.

In the FIF, we selected a  set of Lick-like absorption features (see below), each defined by a central passband flanked by blue and red pseudo-continuum regions. Spectral fitting (see Eq.~\ref{eq:sfit}) is restricted to the wavelength intervals of the index passbands (see~\citealt{NMN:19}). For each index,  both model and observed spectra are normalized by a straight line passing through the corresponding pseudo-continuum bands, using the mean flux values in these bands. 

For the IF, we adopted the same set of absorption features as in the FIF approach and derived abundance ratios by minimizing the following expression:
\begin{equation}
\chi^2 = \sum \frac{ \left( EW_{O} - EW_{M} \right)^2 }{\sigma_{EW}^2},
\label{eq:if}
\end{equation}
where the summation extends over the selected indices. Here, $\rm EW_O$ and $\rm \sigma_{EW}$ are the observed line-strengths (i.e., equivalent widths) and their associated uncertainties, while $\rm EW_M$ denotes the model predictions.

For the FIF and IF methods, we select the following set of spectral indices: the age-sensitive Balmer indices, \hbo , \hgf\ and \hdf ; 
the iron lines Fe4383, Fe4531, Fe5015, Fe5270, and Fe5335; the sodium lines \nai\ and \naii ; the magnesium indices \mgf, \mgb, and \mgi ; the calcium lines \cafr, \cai, \caii, and \caiii ; and additional indices including \atio , \tioi , \tioiir , C4668, 
\cnii , G4300 and Si4101. For most indices, the central passband and pseudo-continua definitions are the same as in the Lick system~\citep{Trager98}, with the exception of \hbo - the optimized \hb\ definition from \citet{CV09}; \cafr , the improved definition of Ca4227 from~\citet{PRS:05}; \tioiir ,  that follows the modified definition of \tioii\ from~\citet{LB:13};  
 \mgf\ and Si4101, defined as in~\citet{Serven:2005}; \atio , defined as in~\citet{Spiniello:2014}; and \naii , defined as in~\citet{CvD12a}, with the modifications described in~\citet{LB:13}~\footnote{
 We also tested the effect of excluding certain indices from the analysis, such as Mg1 (a molecular index), \cai\ (found by~\citealt{SP:2023} to yield biased \cafe\ estimates), and Si4104 (which overlaps significantly with the age-sensitive $\rm H_\delta$ line), resulting in only minor variations, within the quoted error bars.}.

{ For each method, we fit the spectrum with either one or two SSPs. Previous studies showed that the M31 bulge is well described by a single SSP (e.g.,~\citealt{Saglia:2010}), but \citet{Dong:2018} found small fractions of intermediate-age stars, motivating two-SSP models. Massive ETGs are likewise well described by one SSP plus small young-star fractions~\citep{Thomas:2010, SalvadorRusinol:2020}. As shown below, our abundance-ratio estimates are virtually unchanged between one- and two-SSP fits. Although the two-SSP fits assume the same abundance pattern for both components, this assumption does not affect our results. For a population of young age, with intrinsically weak metal lines -- such as the younger SSP resulting from our fits to the M31 bulge (typically $\lesssim$3 Gyr) -- the effect of varying abundance ratios is negligible compared to that for an old ($\gtrsim 10$ Gyr) component.}

Prior to fitting, all models are convolved to match the velocity dispersion $\sigma$ of each observed spectrum. 
To account for contamination from nebular emission in the galaxy spectra, emission lines were modeled as Gaussian profiles and were fit simultaneously to the absorption spectrum in the FSF method. A three-component kinematic model is adopted: one component for the [O III] $\lambda\lambda$4959, 5007 doublet, a second for the [N I] $\lambda\lambda$ 5198, 5200  lines, and a third for the Balmer series (H$\beta$, H$\gamma$, and H$\delta$), each with independent radial velocity and sigma. Additionally, we impose the constraint that the Balmer emission-line strengths decrease monotonically, such that $\rm H\beta > H\gamma > H\delta$. To apply the FIF and IF fitting methods, the best-fitting emission lines from FSF are first subtracted from the input spectra.  

{ In all cases, the best-fitting parameters ($\rm Age_i$, $\rm [Z/H]_i$, $\rm f_i$, \xfe, multiplicative-polynomial coefficients, and emission-line parameters) are obtained by minimizing the $\chi^2$ with a Levenberg–Marquardt algorithm, exploring a grid of initial conditions to ensure convergence to the global minimum. Uncertainties are estimated from Monte Carlo realizations of the input spectrum, perturbing fluxes according to their errors. Using the best-fitting parameters, we compute luminosity-weighted iron and total metallicities (\feh\ and \mh; see App.~\ref{app:mh}) and derive luminosity-weighted ages consistent with~\citet{LB:13} (SDSS stacked spectra) and LB21 (M31 bulge) (see App.~\ref{app:ages}).}

When fitting the spectra with E-MILES models, we adopt bimodal IMF slopes fixed to the values derived by~\citet{LB:13} for the SDSS stacked spectra of ETGs (see their figure~21, case “2SSP+X/Fe”), and by LB21 for the bulge of M31 (see their figure~5, black line). We verified, however, that leaving the IMF slope as a free fitting parameter yields abundance ratio estimates fully consistent with those presented in this work (see Sec.\ref{sec:xfe}). For the fits with CvD models, we treat the low-mass IMF slopes ($\rm x_1$ and $\rm x_2$, see Sec.~\ref{sec:models}) as free parameters. We also verified that adopting a fixed Kroupa-like IMF ($\rm x_1=1.3$, $\rm x_2=2.3$) does not significantly affect our results.

We emphasize that the quality of the fits obtained with both the FSF and FIF methods is excellent, with a mean absolute deviation of the residuals $\lesssim 0.6 \%$ for all M31 spectra, when using either the E-MILES or CvD18 models. In App.~\ref{app:fits}, we illustrate this point by presenting  some examples of best-fitting spectra obtained with the FSF and FIF methods. As for the IF method, we are able to match all M31 indices within the 1--2$\sigma$ confidence level, as already shown in LB21 (see their figures~2 and~3). For the SDSS stacks, all methods reproduce the data with a similar level of accuracy as for the M31 bulge ({see, e.g.,~\citealt{LB:13}}).

\begin{figure*}[]
\begin{center}
 \leavevmode
 \includegraphics[width=17cm]{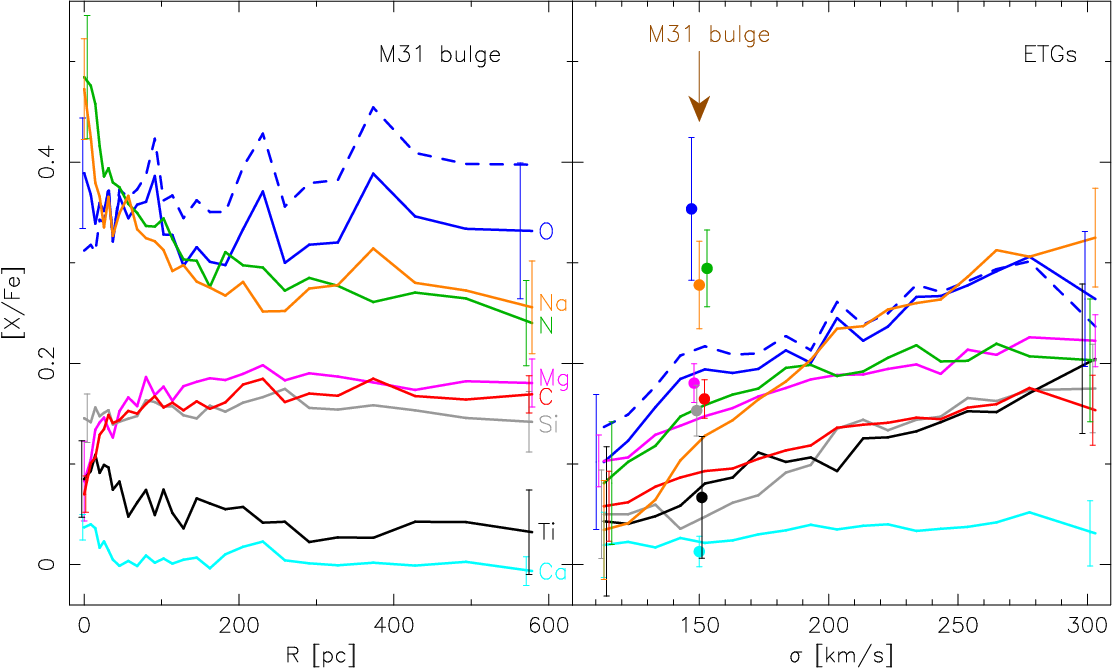}
\end{center}
 \caption{
Median values of abundance ratios, obtained from different models and methods as a function of (left) galactocentric distance, R, for the bulge of M31, and (right) galaxy velocity dispersion for stacked spectra of ETGs. In both plots, different colors refer to different chemical elements (see labels in the left panel). Dashed blue curves show \ofe\ abundances corrected for the Milky Way pattern (see the text). In each panel, error bars are plotted only for the end points. In the right panel, dots with error bars mark abundance estimates obtained by mimicking the Sloan fiber aperture (see the text) for the M31 bulge (see brown arrow). 
For clarity, small horizontal shifts have been applied to the symbols (error bars and dots) corresponding to different elements.
 }
   \label{fig:xfe_rad_sig}
\end{figure*}

\section{Chemical evolution models}
\label{sec:cmodels}
We adopt a revised version of the chemical evolution model { of~\citet{MU:2015}}, which provided predictions for the [$\alpha$/Fe] abundance ratios in the bulge of M31. The model assumes a bulge formation by rapid gas accretion together with an initial strong burst of star formation. 
This is the typical scenario for the formation of a classical bulge and it has been widely adopted in the literature to describe the formation of the Galactic bulge \citep{MB:90,B:07}. However, the MW bulge also hosts a population that may have been accreted from the thin disk, as suggested by kinematic and chemical evidence \citep[see][]{M:19, M:24}. For the bulge of M31, this population becomes significant only at galactocentric distances larger than those probed in the present work~\citep{Diaz:2018}. Therefore, we consider a classical scenario for the formation of the M31 bulge.

The SFR of the model follows a Schmidt-Kennicutt law:
\begin{equation}
\rm
    \psi(t)= \nu \sigma_{gas}^k,
\end{equation}
where $\nu$ (in Gyr$^{-1}$) is the star-formation efficiency, $\sigma_{gas}$ is the gas density, and $\rm k=1.5$.
We assume that the total surface mass density follows a Sersic law with central  density $\rm \sigma_0 =1.28\cdot 10^{5}M_{\odot} pc^{-2}$, and that gas accumulates according to the law:
\begin{equation}
\rm
    \sigma(r,t)= A(r)e^{-t/\tau},
\end{equation}
where $\rm A(r)$ is a parameter fixed to reproduce the present-day total surface mass density, and $\tau$ (in Gyr) is the accretion timescale, which can be constant or vary with galactocentric distance.

Our goal is to investigate radial abundance gradients in the bulge of M31. Gradients in absolute abundances (e.g.,~\feh ) and total metallicity are generally negative -- i.e., metal content decreases with galactocentric distance -- as observed in both spheroids and spiral discs. In contrast, abundance ratio gradients can be negative, positive, or flat, depending on the SFR as a function of galactocentric radius.  In spiral discs, negative metallicity gradients are obtained in an inside-out scenario~\citep[see][]{C:01,BP:99} coupled with a decreasing efficiency of star formation with radius and/or radial gas flows \citep[see][]{P:20}, or by changing the IMF. In spheroidal systems, an outside-in quenching scenario seems to be preferred \citep[see][]{M:98}.

To explore these possibilities, we divide the M31 bulge in three zones: a central one with a radius of 150~pc, and two other zones centered at 300 and 600~pc, respectively, each with a width of 300~pc in galactocentric distance. This spatial segmentation matches the coverage of our data.  We then consider two model scenarios:
\begin{description}
\item Variable IMF -- The initial mass function (IMF) varies with radius, being flatter (i.e., more top-heavy) in the central region than in the outer ones~\footnote{Note that this assumption is not in conflict with LB21, who found a mildly bottom-heavy -- rather than top-heavy -- IMF in the center of the bulge. In fact, LB21 constrained only the low-mass end of the IMF, below $\sim 1$~$\rm M_\odot$, where stars do not contribute significantly to chemical enrichment. Therefore, for simplicity, we did not vary the low-mass end of the IMF in our chemical evolution models.}.  Specifically, we adopt a slope of $x = 1.1$ in the central zone, compared to a Salpeter-like value of $x = 1.35$ elsewhere, in agreement with previous constraints on the high-mass end of the IMF for the M31 bulge \citep[e.g.,][]{B:07}.
In this model, the abundance ratio gradients are relatively flat, with variations in the IMF producing either central enhancements or depressions. These same variations also contribute to gradients in metallicity.
\item Outside-in quenching -- Star formation halts earlier in the outer regions, while it continues longer in the central zone. This scenario produces a decreasing gradient of [$\alpha$/Fe] ratios toward the center, as the longer duration of star formation allows Type Ia supernovae to produce more iron. Meanwhile, both $\alpha$-element and Fe absolute abundances decline with radius. The earlier quenching in the outer regions may naturally result from galactic winds developing first in the outer regions~\citep{M:98}.  Star formation is continuous in the center, although it becomes negligible after $\sim$5~Gyr, while in the intermediate and outer regions, it stops after 600~Myr.
\end{description}

The main parameters of the two models are summarized in Table~\ref{tab_models}.
In the  variable IMF model, we do not stop star formation and assume a very high star-formation efficiency in the center ($\nu$=25$Gyr^{-1}$).
The parameters listed in Table~\ref{tab_models} were chosen ad hoc to reproduce both the MDF  (see App.~\ref{app:MDF}) and the \feh\ radial profile of the bulge (see below), in agreement with a classical bulge formation scenario.

\begin{table*}
\tiny
\caption{Summary of the main parameters of the chemical evolution models for the M31 bulge.}
\label{tab_models}
\begin{tabular}{|c|cccc|cccc|cccc|}
\hline\hline
 & \multicolumn{4}{|c|}{0 pc} & \multicolumn{4}{|c|}{300 pc} & \multicolumn{4}{|c|}{600 pc} \\
Models & $\tau$ & $\nu$ & SFR=0 & IMF & $\tau$ & $\nu$ & SFR=0 & IMF & $\tau$ & $\nu$ & SFR=0 & IMF \\
 & [Gyr] & [Gyr$^{-1}$] &  &  & [Gyr] & [Gyr$^{-1}$] &  &  & [Gyr] & [Gyr$^{-1}$] &  &  \\
\hline 
 & \multicolumn{4}{|c|}{}&  \multicolumn{4}{|c|}{}& \multicolumn{4}{|c|}{}\\
VARIABLE IMF &0.1&25.0&/&x=1.1&0.1&25.0&/&x=1.35&0.1&25.0&/&x=1.35\\ 
 & \multicolumn{4}{|c|}{}&  \multicolumn{4}{|c|}{}& \multicolumn{4}{|c|}{}\\
\hline
& \multicolumn{4}{|c|}{}&  \multicolumn{4}{|c|}{}& 
\multicolumn{4}{|c|}{}\\
OUTSIDE-IN&0.1&5.0&/&x=1.35&0.1&25.0&$t>0.6$ Gyr&x=1.35&0.1&25.0&$t>0.6$ Gyr&x=1.35\\ 
 & \multicolumn{4}{|c|}{}&  \multicolumn{4}{|c|}{}& \multicolumn{4}{|c|}{}\\
 \hline
\end{tabular}
\tablefoot{
 We report, for each model and radial bin, the gas infall timescale ($\tau$), star-formation efficiency ($\nu$), any imposed cessation of star formation ("SFR=0"), and the IMF slope $\rm x$. }
\end{table*}

We follow the evolution of several chemical species, including C, N, O, Mg, Si, Ca, Ti, Na and Fe. Among these, O, Mg, Si, Ca and Ti are $\alpha$-elements.
Concerning the stellar yields -- which are a critical ingredient of chemical evolution models -- we adopt two different sources: one set refers to yields taken from the literature~\citep[hereafter R10]{Romano:10}, while the other one is from~\citet[hereafter F04]{Francois:04}, including modifications for specific elements to match solar-neighborhood observations.
We also account for element production by Type Ia SNe and low and intermediate mass stars (LIMS). In particular, Type Ia SNe are the major producers of Fe, while LIMS are the major producers of N. However, a small fraction of N is also produced in massive stars. While most of the N from LIMS is secondary (proportional to the initial stellar metallicity), N from massive stars should be of primary origin \citep{M:86,MM:02,LC:18}.\\
We validated our assumptions by reproducing the metallicity distribution function (MDF) of M31 bulge stars observed { by~\citet{SJ:2005}} (see App.~\ref{app:MDF}).\\
To compare our results with observations, we computed average elemental abundances weighted over stellar mass \citep[see][]{M:01}. Although observational determinations are luminosity-weighted, it has been shown that, for systems more massive than $\rm 10^{9}M_{\odot}$, mass- and luminosity-weighted abundances coincide   \citep{MPG:98,M:01}.

In particular,  the mean abundance ratio of an element X relative to Fe is computed as:
\begin{equation}
\rm
\langle \mathrm{X/Fe} \rangle = \frac{1}{S_f} \int_0^{S_f} (\mathrm{X/Fe})(S) \, \mathrm{d}S,
\end{equation}
where $(\mathrm{X/Fe})(S)$ denotes the abundance ratio of the interstellar medium out of  which  stars form; $S_f$ is the total mass of stars ever born up to the present time contributing to the mass, while $\rm S=S(t)$ is the mass of stars born until the time $\rm t$.
After computing the average, we converted it into logarithmic form to obtain the bracket notation:
\begin{equation}
[\mathrm{<X/Fe>}] = \log \left( \langle \mathrm{X/Fe} \rangle \right) - \log(\mathrm{X/Fe})_\odot,
\end{equation}
as described in \citet{G:96}. A similar approach is used for calculating $[\mathrm{Fe/H}]$. The assumed solar abundances are those of \citet{Asplund:2009}.

\section{Results}
\label{sec:results}

\subsection{Metallicity}
\label{sec:met}
Figure~\ref{fig:xh_m31} shows the radial metallicity profiles of the M31 bulge, i.e., \feh\ (top panel) and \mh\ (bottom panel), respectively. The iron metallicity is directly derived from our fitting procedure, while \mh\ is estimated from \feh\ and the inferred abundance ratios, as detailed in App.~\ref{app:mh}. Both profiles flatten at radii $\gtrsim 100$~pc , where \feh\ is $\sim -0.2$~dex, and \mh\ is mildly supersolar ($\sim 0.1$~dex). Toward the center, both iron and total metallicity rise steeply, peaking at $\sim 0.2$~dex and $\sim 0.5$~dex, respectively. 

{ \citet{SJ:2005}} derived the MDF of RGB stars in the M31 bulge, finding a peak at $\mh \sim 0$ (see App.~\ref{app:MDF}). While this value is lower than our estimate at $\rm R \gtrsim 100$~pc -- representative of the bulk of the bulge -- it is important to note { that~\citet{SJ:2005}} used solar-scaled isochrones to fit the observed color-magnitude diagram of RGB stars, and subsequently applied a correction factor to account for $\alpha$-enhancement (see App.~\ref{app:mh}), to convert \feh\ into \mh . Assuming all $\alpha$-elements vary in lockstep with an enhancement of $\sim 0.3$~dex, they applied a correction corresponding to a shift of $\sim 0.2$~dex in metallicity. Removing this shift places the peak of { the distribution} at \feh$ \sim -0.2$~dex, fully consistent with our inferred values at $\rm R \gtrsim 100$~pc . Interestingly, previous studies of unresolved stellar populations ({ e.g.,~\citealt{Saglia:2010, Saglia:2018}}, LB21) found the bulk of the bulge to have approximately  solar metallicity. However, these results were obtained either by assuming all $\alpha$-elements vary in lockstep, or by estimating \mh\ from Mg and Fe absorption features, which do not capture the full abundance pattern.

Figure~\ref{fig:xh_sdss} plots the metallicity of ETGs, derived from the SDSS stacked spectra, as a function of stellar velocity dispersion, $\sigma$. We find that \feh\ varies only mildly from $\sim -0.1$~dex at $\sigma \sim 110$~\kms\ to $\sim 0.07$~dex at $\sigma \sim 300$~\kms . This result is consistent with previous findings ({ e.g.,~\citealt{CGvD:2014}}). On the other hand, \mh\ shows a larger variation, increasing from  solar value at lowest $\sigma$ to $\sim 0.3$~dex for the most massive ETGs. At the velocity dispersion typical of the M31 bulge ($\sigma \sim 150$~\kms ; see vertical gray arrows in the Figure), ETGs exhibit typical metallicities of \feh$\sim -0.06$~dex and \mh$\sim 0.08$~dex. 

To compare the metallicity content of ETGs with that of the M31 bulge, we estimate the bulge metallicities by mimicking the effect of the SDSS 1.5\arcsec -radius fiber aperture. To this effect, we consider low- ($\sim 150$~\kms ) and high- ($\gtrsim 260$~\kms ) $\sigma$ ETGs, for which the SDSS fiber samples a relative aperture of $\rm A_{SDSS}=1.5/R_e \sim 0.6$ and $0.25$, respectively~\footnote{These estimates are based on structural parameter estimates for SPIDER ETGs from LB10. Selecting galaxies with reliable structural parameters (i.e., relative uncertainty on $\rm R_e$ less than 50$\%$),  we estimate typical effective radii of $\sim 2.5$~\arcsec\ and $\sim 6$~\arcsec\ for low- and high-$\sigma$ ETGs, respectively.}. For both \feh\ and \mh , we compute the luminosity-weighted mean metallicity of the bulge within $\rm A_{SDSS}$, by integrating the corresponding profiles (Figure~\ref{fig:xh_m31}). The \feh\ and \mh\ values at each radius are weighted by the luminosity enclosed in an annular bin centered at that radius, assuming circular symmetry. The luminosity is computed by using a single-S{\'e}rsic surface brightness profile for the bulge, with Sersic index $\rm n=2.2$ and effective radius of 1~kpc~\citep{Courteau:2011}. The resulting M31 metallicities, corresponding to the apertures of low- and high-$\sigma$ galaxies, are plotted in Fig.~\ref{fig:xh_sdss} as gray triangles with error bars. At low-$\sigma$, we obtain average differences of $-0.08 \pm 0.04$~dex in \feh\ and $0.05 \pm 0.07$~dex  in \mh\ between M31 and ETGs. At high-$\sigma$, the corresponding differences are  $-0.14 \pm 0.03$~dex and $-0.11 \pm 0.06$~dex, respectively. We conclude that the bulge of M31 has significantly lower metallicity than most massive ETGs. Compared to low-mass ETGs, the bulge has slightly lower \feh\ (by $\sim 0.1$~dex), but similar \mh . This result is due to the detailed abundance pattern of the M31 bulge, as discussed below. 

\subsection{Abundance ratios}
\label{sec:xfe}
Figure~\ref{fig:xfe_m31} presents the individual abundance ratios (in separate panels) for the bulge of M31, derived with different methods (Sect.~\ref{sec:methods}). Results are summarized as follows.

{ Average differences between 2SSP and 1SSP fits are $<0.05$~dex for all elements. Across the FSF, FIF and IF methods, mean differences are also within $0.05$~dex, except  for N (IF gives values $\sim 0.06$~dex lower than the median) and Ti (FSF gives values $\sim 0.07$~dex higher). For Mg, Ca, and C, \xfe\ estimates are highly consistent across all methods, reflecting the strong sensitivity of specific spectral features to these elements -- the \mgb\ triplet for Mg, the Ca triplet and Ca4227 lines for Ca, and the C4668 index for C. In contrast,  O and Ti show larger scatter, with differences up to $\sim 0.2$~dex. The large \ofe\ scatter  arises from the lack of strong atomic features directly tracing oxygen: at $\lambda < $5800~\AA\ changes in \ofe\  mainly affect the spectrum through the effect on molecular dissociation equilibrium, while at larger wavelengths the effect is on molecular lines (e.g., TiO; {see~\citealt{CGvD:2014}}). 
The trends for the M31 bulge are consistent with those for SDSS stacked spectra (App.~\ref{app:sdss}). As discussed in App.~\ref{app:sims}, tests on mock spectra, with known abundance patterns, confirm that all methods recover the input \xfe\ with negligible bias. Thus, variations among methods in Fig.~\ref{fig:xfe_m31} (and Figure.~\ref{fig:sdss_xfe}) are likely due to (percent-level) systematic deviations between models and data arising from observational or model uncertainties.}

Most radial profiles are symmetric around the photometric center of the bulge. A small asymmetry (at the few percent level) is observed only for Ti , with slightly higher \tife\ values at negative $\rm R$. Nitrogen and sodium exhibit negative radial gradients, peaking at the center of the bulge, while magnesium and carbon show the opposite behavior, with a central dip at galactocentric distances $\lesssim$50--100~pc . The remaining elements -- O, Si, Ca, and Ti -- show flat radial trends, with no significant gradients. 

These results are further illustrated in Figure~\ref{fig:xfe_rad_sig}, which shows the median trends of \xfe\ as a function of galactocentric distance, $\rm R$, in the bulge of M31 (left panel), and stellar velocity dispersion, $\sigma$, in SDSS ETGs (right panel). For clarity, error bars~\footnote{Note that error bars reflect the dispersion across methods, which dominates over the statistical uncertainties from the Monte-Carlo simulations (see Sec.~\ref{sec:methods}). } are shown only for the innermost and outermost radial bins in the bulge of M31, and only for the  lowest and highest $\sigma$ in the SDSS stacked spectra.
For the bulge of M31, the median trends are obtained by mirroring and interpolating the \xfe\ values from the negative side of the slit to the positive-$\rm R$ side, and taking mean values. Uncertainties reflect the standard deviation of estimates from different methods, which dominate over formal errors. As mentioned in Sec.~\ref{sec:methods}, the \ofe\ values should be corrected for the MW abundance pattern, imprinted in the stars used to construct the empirical SSP models (either E-MILES or CvD18). To perform the correction, we model the trend of MW disk stars in figure~15 of~\citet{Bensby:2014} with a linear relation: $\rm [O/Fe]_{MW} = -0.38 \cdot [Fe/H]_{MW}$. For each spectrum, with given \feh , we computed the corresponding $\rm [O/Fe]_{MW}$, and then added this up to the \ofe\ estimate from our fitting procedures. The corrected \ofe\ trends are shown with blue dashed lines in Fig.~\ref{fig:xfe_rad_sig}. The correction does not affect significantly the trend of \ofe\ with $\sigma$ for ETGs, while it tends to flatten, or even invert (with slightly smaller values in the center), the trend of \ofe\ with $\rm R$ for the M31 bulge. Consistent with what seen in Figure~\ref{fig:xfe_m31}, Figure~\ref{fig:xfe_rad_sig} shows that O, Na, and N have the highest enhancements, with \xfe\ values up to $\sim 0.5$~dex in the center. Nitrogen and sodium increase steeply towards the center, while oxygen displays a flatter profile~\footnote{However, we caution that in the very central region of the bulge ($\lesssim 20$~pc), where the metallicity reaches values as high as \mh$\sim 0.5$~dex, the abundance ratio estimates are more uncertain, owing to limitations in the response functions at such high metallicity. }.
 Carbon and magnesium closely track each other, declining towards the center to values $\lesssim$0.1~dex. Silicon abundance remains nearly constant across  the bulge~\footnote{However, we caution the reader that in the optical spectral range, there is no individual feature with prominent sensitivity to the Si abundance. }, at levels comparable to C and Mg. Titanium shows lower values, with a marginal trend to increase at small $\rm R$. However, \tife\ estimates are significantly dependent on the adopted methodology, as discussed above. Calcium, in contrast, follows iron closely, with \cafe$\sim 0$.  

The comparison with SDSS ETGs reveals striking differences. At the characteristic velocity dispersion of the bulge ($\sigma \sim 150$~\kms ), the abundance ratios in ETGs are typically below $\sim 0.2$~dex for all elements (see the brown vertical arrow in Figure~\ref{fig:xfe_rad_sig}), in stark contrast to the bulge, where elements such as N, Na, and O, are strongly enhanced. Overall, the abundance pattern of the bulge more closely resembles that of very massive ETGs ($\sigma \sim 280$~\kms), with the exception of Ti, which is significantly enhanced in massive galaxies but remains $\lesssim 0.1$~dex in the bulge of M31 (albeit with large error bars). In order to perform a more direct comparison, we adopted the same approach as above (see Sec.~\ref{sec:met}), estimating the abundance ratios of the M31 bulge by mimicking the effect of the SDSS 1.5''-radius fiber aperture. In particular, the filled dots with error bars in~\ref{fig:xfe_rad_sig} (right panel) show the M31 abundance ratios corresponding to SDSS ETGs with the same $\sigma$ as the bulge ($ \sim 150$~\kms ). The M31 bulge is significantly more enhanced than low-$\sigma$ ETGs.

\begin{figure}[]
\begin{center}
 \leavevmode
 \includegraphics[width=7.5cm]{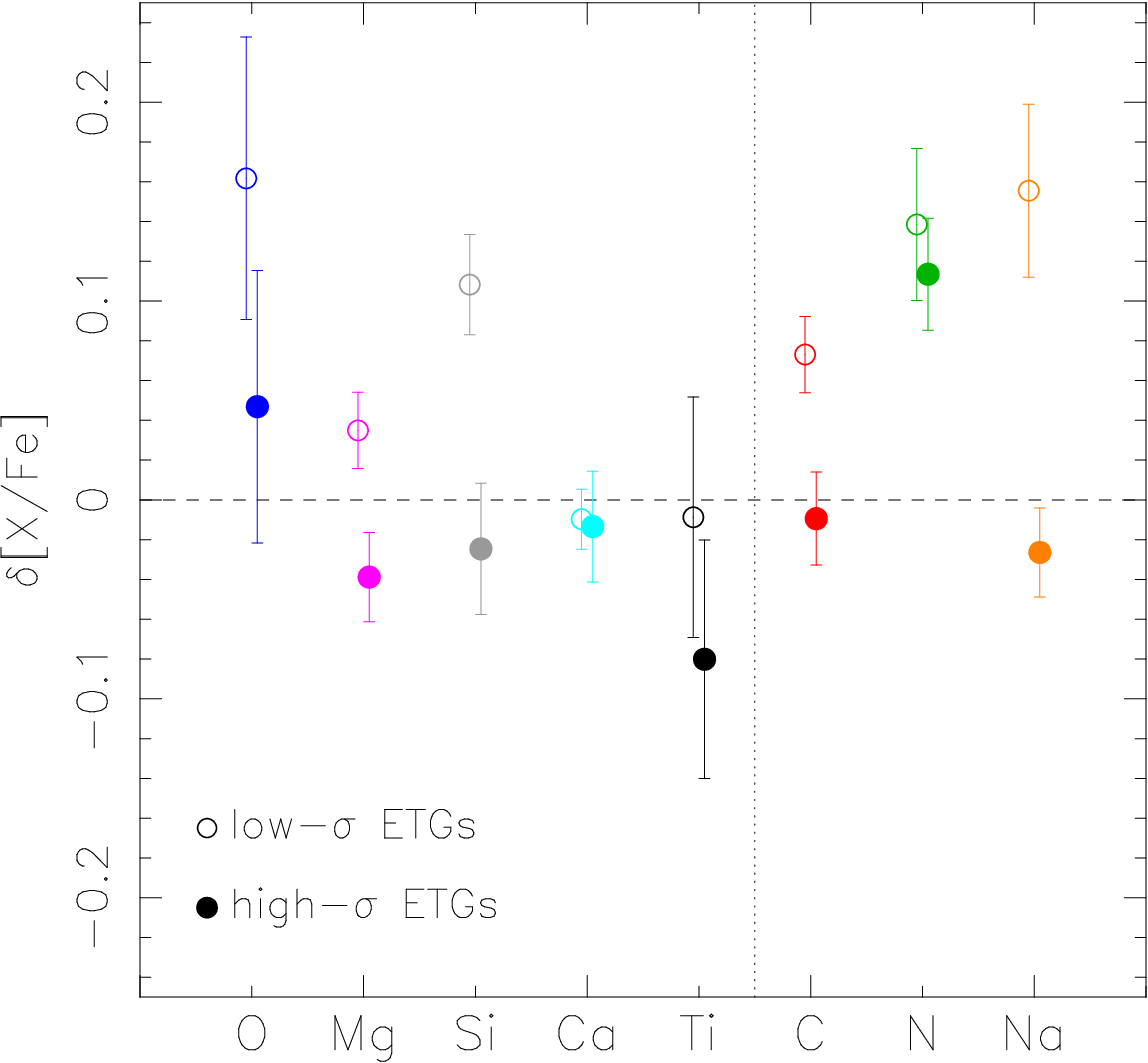}
\end{center}
 \caption{
 Median difference in abundance ratios between the bulge of M31 and stacked spectra of ETGs for different elements, as shown on the x-axis. Median values and error bars are derived from different models and fitting methods (see the text). Different elements are color-coded as in Fig.~\ref{fig:xfe_rad_sig}. Empty and filled symbols denote ETGs with low- ($\sigma = 150$~\kms ) and high- ($\sigma > 260$~kms ) velocity dispersions, respectively, as indicated in the lower-left corner. The horizontal dashed line marks a difference of zero, while the vertical dotted line separates the $\alpha$-elements (left) from the remaining ones (right). 
 }
   \label{fig:dxfe_m31_etgs}
\end{figure}

Figure~\ref{fig:dxfe_m31_etgs} shows the average difference, $\rm \delta [X/Fe]$, between the SDSS aperture-corrected abundance ratios of the M31 bulge and those of low- ($ \sim 150$~\kms ) and high- ($ \sim 280$~\kms ) $\sigma$ ETGs, plotted as empty and filled symbols, respectively. 
The differences are averaged over the various models and fitting methods.
As anticipated from Fig.~\ref{fig:xfe_rad_sig}, the abundance pattern of the bulge is more similar to that of high-$\sigma$ ETGs: for low-$\sigma$ systems, $\rm \delta [X/Fe]$ is systematically positive for most elements (in particular, O, Si, C, N, Na), while for high-$\sigma$ ETGs, the differences are consistent with zero for all elements except N, which is further enhanced in the M31 bulge by $\sim 0.1$~dex. These findings also explain the differences in metallicity content between the M31 bulge and ETGs (Fig.~\ref{fig:xh_sdss}):
the bulge has lower \feh\ than ETGs with the same $\sigma$,  but comparable \mh , owing to stronger abundance enhancements, that are similar to those of the most massive ETGs. 

\begin{figure*}[]
\begin{center}
 \leavevmode
 \includegraphics[width=18cm]{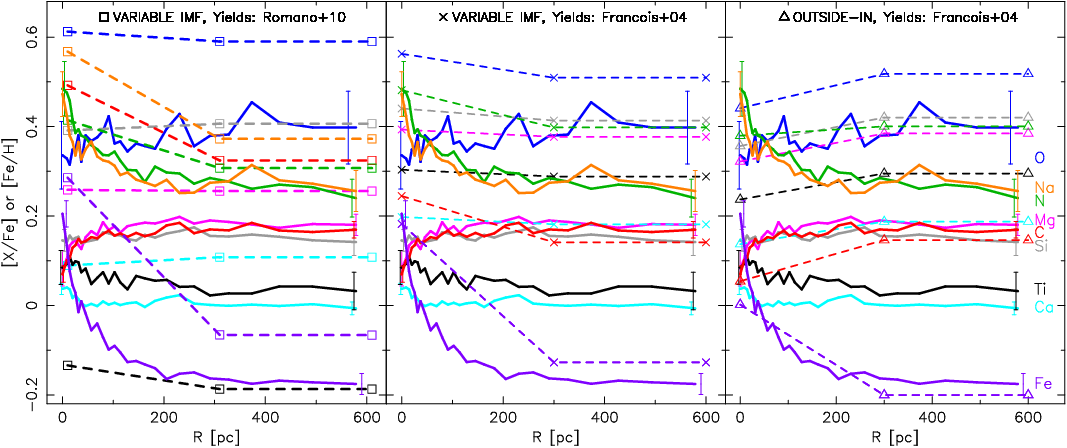}
\end{center}
 \caption{Comparison of abundance ratios, \xfe , and metallicity, \feh ,  as a function of galactocentric distance for the bulge of M31 (solid lines) with predictions from our chemical evolution models (dashed lines).
The left and middle panels show models for a variable IMF scenario, adopting stellar yields from R10 and F04, respectively. The right panel shows models for an outside-in quenching scenario, with stellar yields from F04.
All models are computed for the center and for two additional regions at radii of 300 and 600~pc (empty squares, crosses, and empty triangles in the left, middle, and right panels, respectively).
Different elemental abundances and \feh\ are  plotted in various colors (see labels in the rightmost panel). Error bars on observed data points are shown only for the innermost and outermost bins, as in Fig.~\ref{fig:xfe_rad_sig}.
We corrected oxygen abundances for the Milky Way pattern, as detailed in Sec.~\ref{sec:xfe}.
 }
   \label{fig:xfe_chem}
\end{figure*}

\subsection{Comparison to chemical evolution models}
\label{sec:res_cmodels}

{ Figure~\ref{fig:xfe_chem} compares the radial trends of abundance ratios and \feh\ in the M31 bulge with predictions from our chemical evolution models (Sec.~\ref{sec:cmodels}) for a variable IMF scenario using R10 and F04 yields (left and middle panels) and for an outside-in quenching scenario with F04 yields (right panel). Models are shown for the bulge center and two additional regions at 300 and 600~pc.

In the intermediate and outer regions, the models’ rapid star formation produces flat radial abundance trends, consistent with the data. The resulting MDFs match observations of individual bulge { stars~\citep{SJ:2005}}, as shown in Appendix~\ref{app:MDF}, which also presents the time evolution of elemental abundances in the F04 models. Element-specific discrepancies between models and data depend on the adopted stellar yields, as illustrated by the comparison of the left and middle panels:}
\begin{description}
\item Metallicity -- Both R10 and F04 models predict slightly subsolar \feh . F04 yields provide a slightly better match to the data (within $\sim 0.05$~dex). 
\item Oxygen and Magnesium -- All models predict a significant enhancement of the light $\alpha$-elements, with \ofe $>$ \mgfe , consistent with the data. For oxygen,  the F04 yields provide a closer match to the observations, whereas for magnesium, the R10 models perform better (within $\sim 0.1$~dex).
\item Other $\alpha$ elements --  \sife\ is significantly overestimated in the models (by $>0.2$~dex), suggesting potential issues either with the adopted yields or with the spectral response functions used to infer \sife\ (see Sec.~\ref{sec:models}). \tife\ is underpredicted by $\sim 0.2$~dex with R10 yields, while being overpredicted by a similar amount with F04, highlighting the uncertainties in predicting  Ti abundance. \cafe\ is better reproduced by the R10 model, although it remains slightly overestimated compared to the data (by $\sim 0.1$~dex). This may suggest a larger contribution of Ca from Type Ia SNe than currently assumed~\citep{Kobayashi:2020}. 
\item Carbon and Nitrogen -- \nfe\ is well reproduced by the R10 model (within $0.05$~dex), while it is overpredicted (by $\sim 0.15$~dex) with the F04 model, which also includes the primary production of nitrogen from massive stars. While such production is required by independent observations of Galactic halo stars and damped Lyman-$\alpha$ systems~\citep{M:86, MMV:97}, the overprediction here suggests that nitrogen yields should be further refined, and/or that \nfe\ may be underestimated in the data (due, e.g., to the adopted response functions). \cfe\ is well matched by the F04 model, while R10 slightly overpredicts it.
\item Sodium -- \nafe\ estimates are only available for the R10 model, showing good agreement (within $\sim 0.1$~dex) with the data. Previous estimates from LB21 reported \nafe\ values $\sim 0.1$~dex higher than in this work, implying even better agreement with our chemical evolution model. 
\end{description}

In the intermediate and outer regions, both the outside-in and variable IMF scenarios produce nearly identical results when the same set of stellar yields is used (i.e., F04), as expected given the similar model parameters in these regions (see Tab.~\ref{tab_models}).
In the central region, however, the two formation scenarios diverge:
\begin{description}
\item -- The variable IMF model predicts an increase in \xfe\ toward the center due to a flatter IMF slope favoring massive stars;
\item -- The outside-in scenario predicts a decrease in \xfe\ in the center, as extended star formation allows more iron to accumulate from Type Ia SNe.
\end{description}
Interestingly, each model predicts a uniform radial behavior for all $\alpha$-elements, in contrast with the observational data, where we observe a decoupled radial trend among different elements (Sec.~\ref{sec:xfe}): \cfe , \mgfe\  and, to a lesser extent, \ofe\ decrease toward the center, while \tife , \cafe , and \sife\  remain relatively flat across the bulge. This decoupling is difficult to explain unless some physical process affects different chemical elements in distinct ways, as further discussed in the following section. 

\section{Discussion}
\label{sec:discussion}

\subsection{Massive ETGs and the M31 bulge}
\label{sec:disc_etgs}
We find that the abundance ratios of $\alpha$-elements in ETGs increase with galaxy mass (see left panel of Fig.~\ref{fig:xfe_rad_sig}),  consistent with previous studies ({ e.g.,~\citealt{Thomas:2005, Bernardi:2006, Thomas:2010, Smith:2009, JTM:2012, CGvD:2014, Parikh:2019, Li:25}}). This trend is commonly attributed to a mass-dependent star-formation timescale, with more massive systems forming their stars more rapidly.  In particular, \citet{M:94} proposed that massive ellipticals experience higher star-formation efficiency, leading to older stellar populations and shorter star-formation durations ($\leq 1$ Gyr). 
Since light $\alpha$-elements, such as O and Mg, are only produced by core-collapse SNe from short-lived massive stars, while Fe is also synthesized by Type Ia SNe from longer-lived low-mass progenitors, a shorter star-formation timescale naturally leads to enhanced \mgfe\ and \ofe\ in high-mass systems. Alternatively, the enhancement of $\alpha$-elements may reflect a top-heavy IMF during the early phases of formation, increasing the metallicity and $\alpha$-elements content in massive galaxies~\citep{Weidner:2013, F:15, Fontanot:18, Jerabkova:2018}. 
Indeed, as seen in Fig.~\ref{fig:xfe_rad_sig}, not all the $\alpha$ elements exhibit the same trend with galaxy mass. The elements with high atomic number (Si, Ti, and in particular Ca) show weaker dependence on $\sigma$, likely due to their additional production by Type Ia SNe~\citep{Nomoto:2013}, which reduces the difference in \xfe\ between low- and high-mass systems.  
In this context, the finding that the bulk of the M31 bulge exhibits $\alpha$-element abundance ratios comparable to those of the highest-$\sigma$ ETGs supports a scenario where it formed through a similarly fast and intense star-formation episode as most massive galaxies. This interpretation is consistent with the fact that our study probes the innermost $\lesssim 0.6$kpc, a region dominated in light by a classical bulge component, as shown by~\citet{Bagna:2017}. 

\subsection{Radial behavior of different elements in the M31 bulge}

Our analysis reveals a relative enhancement of the light $\alpha$-elements O and Mg, with  \omg $\gtrsim 0.15$~dex, in both the M31 bulge and massive ETGs. This result may seem surprising, as it contrasts with observations in both the bulge and the disk of the MW, where \omg\ is found to decrease with increasing metallicity, reaching subsolar values at high \feh ~\citep{Bensby:2004, Bensby:2005, Bensby:2013}. This trend has been interpreted as the result of metallicity-dependent stellar winds in massive stars, which lead to significant mass loss and reduced O yields, with a minimum impact on Mg~\citep{McWilliam:2008}. Remarkably, our chemical evolution models not only reproduce the observed MDF of individual bulge stars, but also account for the relative enhancement of O with respect to Mg. This agreement suggests that such high \omg\ ratios may be naturally explained in systems that experienced short and intense star-formation episodes at near-solar metallicity, where oxygen yields are not yet significantly reduced by stellar winds.

We find that  different $\alpha$-elements exhibit distinct radial behaviors in the M31 bulge. In particular, magnesium decreases toward the central 10'', in agreement with previous studies ({ \citealt{Davidge:1997, Saglia:2010}}). If \mgfe\ traces the star-formation timescale, this decline implies a more prolonged star formation in the bulge center. Alternatively, the inner bulge may have accreted stellar material with lower \mgfe , either from the disk by secular evolution processes or from external systems via merging events ({ \citealt{Saglia:2010, Dong:2015, Dong:2018}}).
The interpretation of the \ofe\ profile is less straightforward. After correcting for the MW pattern, our \ofe\ estimates appear to remain flat -- or even decrease -- toward the center, showing a qualitative similarity to the Mg trend. However, in the super-solar metallicity regime of the central bulge, the determination of abundance ratios for some elements, such as O, may be more affected by systematic uncertainties, hampering the interpretation of the radial trends.
Interestingly, the C abundance also follows the Mg trend, decreasing toward the bulge center, possibly indicating a common origin for these elements. A similar conclusion was reached by~\citet{Parikh:2019}, who studied radial variations of C and Mg abundances in ETGs.
We emphasize that Mg and C are particularly well constrained in our analysis, owing to the strong sensitivity of specific spectral features to their abundance ratios (see App.\ref{app:sims}). Moreover, our results are consistent with those of~\citet{Romano:2020}, who found evidence that C is predominantly produced by rotating massive stars.

Among our chemical enrichment models, the one that  best reproduces the radial trends of Mg and C -- the most reliably constrained abundances -- as well as \ofe, is the model assuming an outside-in quenching scenario. This scenario is compatible with the properties of spheroidal systems, as shown by several previous studies~\citep{M:98,P:08}. In this framework, the central decrease in abundance ratios arises from a more prolonged star-formation period in the inner regions, allowing Type Ia SNe more time to pollute the ISM with Fe, thereby reducing the \xfe\ ratios toward the center. The fact that star formation stops earlier in the outer regions can be explained by the onset of a gas outflow,  which inhibits star formation and naturally proceeds outside-in. However, this scenario would, in principle, predict similar behavior for all $\alpha$ elements. The fact that the heavier $\alpha$-elements -- Si, Ti, and Ca -- do not exhibit a central gradient supports the idea that these elements are also significantly produced by Type Ia SNe (see Sec.~\ref{sec:disc_etgs}), making their relative abundance less sensitive to variations in the star-formation timescale. 
Moreover, our chemical models do not account for the effect of radial gas flows, which could occur if the bulge formed through gas infall. Differential radial flows could actually produce the observed decoupled radial trends, although invoking such a mechanism remains difficult to justify on physical grounds.

On the other hand, the high Fe content observed in the center of the M31 bulge could also be explained by a scenario where the high-mass end of the IMF varies with galactocentric distance, favoring the formation of high-mass stars toward the center. LB21 found that the low-mass end of the IMF becomes more bottom-heavy toward the bulge center. This IMF gradient is qualitatively similar -- albeit shallower -- to that observed in massive ETGs~\citep{NMN:15, vanDokkum:2017, LB:19}. As proposed by~\citet{Weidner:2013, F:15, Fontanot:18}, variations at the low and high-mass ends of the IMF might be actually coupled: the explosion of a large number of SNe might eject enough energy into the ISM to promote fragmentation at low-mass scales (see also~\citealt{DeMasi:19}). In this context, the results of LB21 may support chemical enrichment in a variable IMF scenario. However, as shown in Sec.~\ref{sec:res_cmodels}, such a scenario fails to reproduce the central decrease in Mg and C, although it provides a good match to the radial behavior of other elements, such as, in particular, the central enhancement of N and Na. One possible way to explain the central decrease of Mg, C, and O, within a variable IMF scenario is to invoke differential galactic winds. Since galactic winds are primarily driven by core-collapse SNe -- which explode in clustered environments -- the expelled material tends to be rich in $\alpha$-elements and Fe produced by massive stars. Therefore, such winds would preferentially remove the $\alpha$-elements and Fe formed in massive stars.
This may be consistent with the results of~\citet{Liu:2010}, who found a low $\rm O/Fe$ abundance in the diffuse hot gas at the center of the  M31 bulge, suggesting that oxygen might have been significantly depleted from the gas.
Differential winds may also help to explain the high \nfe\ in the bulge center. The outside-in scenario alone does not reproduce the central rise in N abundance. This may be due to uncertainties in its nucleosynthesis: nitrogen is produced both as a primary and secondary element, but its yields remain rather uncertain. Differential galactic winds would not affect the abundance of N, as this element is mainly produced in LIMS, which die as white dwarfs and can form in isolation. This argument has also been invoked to explain the high N abundances observed in high-redshift galaxies with JWST~\citep[see][]{R:25}.

In summary, our analysis shows that both a variable IMF and an outside-in quenching scenario are able to reproduce the abundance pattern of the bulk of the bulge, together with its high central metallicity. However, differential galactic winds appear to be required in order to reproduce the decoupled radial behavior in the abundance ratios of individual chemical elements.

\begin{acknowledgements}
{ We thank the anonymous referee for helpful comments and John
Beckman for constructive discussions.  F.L.B., A.P., and E.E. acknowledge
support from INAF minigrant 1.05.23.04.01; F.M. from INAF Theory Grant
1.05.12.06.05; E.S.  and F.M.  from INAF minigrant 1.05.24.07.02.
F.L.B.  also acknowledges support from Fundaci{\'o}n Occident and the
Instituto de Astrof{\'i}sica de Canarias under the 2022–2025 Visiting
Researcher Program.  AV acknowledges support from
PID2021-123313NA-I00 and PID2022-140869NB-I00 (Spanish Ministry of
Science and Innovation).  This work was partially supported by the IAC
project TRACES, funded by the state and regional budgets of the Canary
Islands Autonomous Community.}
\end{acknowledgements}

\begin{appendix} 

\section{Recovering abundance ratios: Simulations}
\label{app:sims}

\begin{figure*}
\begin{center}
 \leavevmode
 \includegraphics[width=16cm]{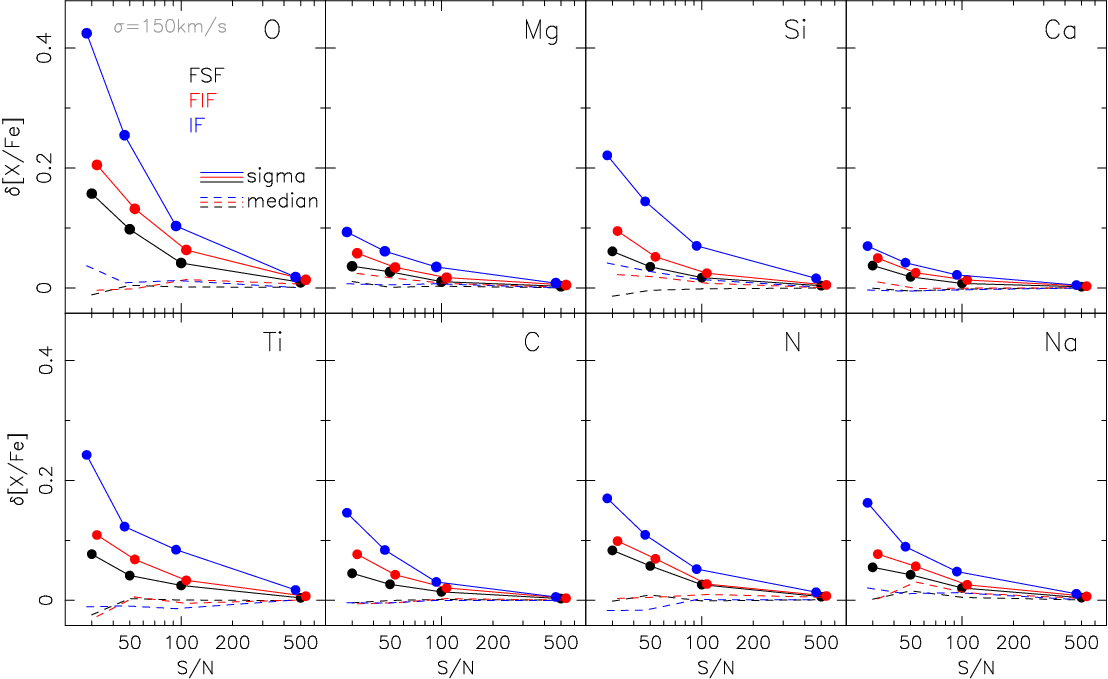}
\end{center}
 \caption{
Results of simulations for low-mass ETGs ($\sigma = 150$~\kms ; see the text), showing the ability to recover individual abundance ratios as a function of S/N ratio, using different techniques, i.e., full spectral fitting (black), full-index fitting (red), and index fitting (blue), as labeled in the top--left panel. In each panel, solid lines with filled circles represent the 1--sigma uncertainties on abundance ratios, while dashed lines show the average difference between recovered and input values. Each panel corresponds to a different element, arranged in the same order as in Fig.~\ref{fig:xfe_m31}. 
 }
   \label{fig:sims_ls}
\end{figure*}

\begin{figure*}
\begin{center}
 \leavevmode
 \includegraphics[width=16cm]{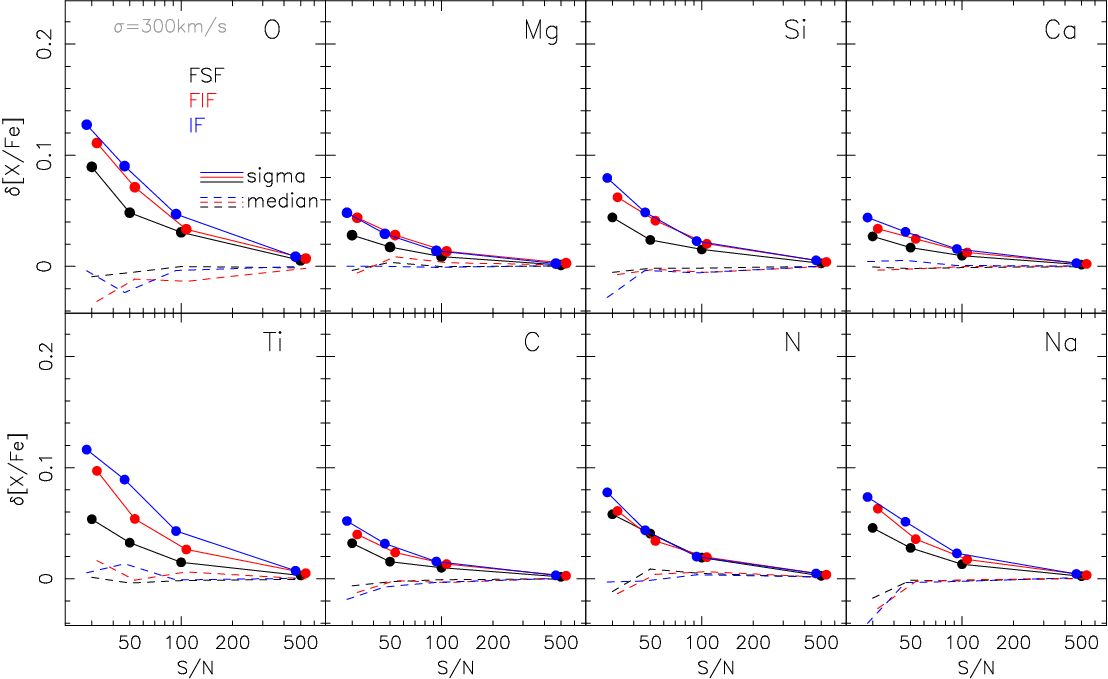}
\end{center}
 \caption{Same as Fig.~\ref{fig:sims_ls} but for simulations of high-$\sigma$ ETGs' spectra (300~\kms ; see the text). 
 }
   \label{fig:sims_hs}
\end{figure*}

We assess the ability of different fitting methods (see Sect.~\ref{sec:methods}) to recover elemental abundance ratios by simulating galaxy spectra with abundance patterns representative of low- and high-$\sigma$ ETGs.  The simulations are based on E-MILES BaSTI SSP models, modified by applying response functions from the CvD18 models to reproduce the desired abundance patterns. Since in the present work we are not interested in constraining the IMF, we adopt a fixed, bimodal, IMF, with a slope of \gammab$=1.3$ (consistent with a Kroupa-like distribution).
 For low (high) $\sigma$ galaxies, we assume an abundance pattern with \xfe$=0.1, 0.1, 0.05, 0, 0$ (0.3, 0.3, 0.2, 0, 0.1), for the $\alpha$-elements O, Mg, Si, Ca, and Ti, respectively, and \xfe$=0.05, 0.1, 0.1$ (0.2, 0.3, 0.3), for C, N, and Na, respectively.
For low-$\sigma$ galaxies, we simulate spectra by combining two SSPs: an old component with an age of 13~Gyr, and a younger SSP, contributing 10$\%$ of the total mass, with an age of 1~Gyr. Both components have a subsolar metallicity of \zh$=-0.1$, and the spectra are broadened to $\sigma$ of $150$~\kms . For high-$\sigma$ galaxies, we simulate spectra using a single, old SSP (13~Gyr) with solar metallicity ( \zh$=0.06$) and $\sigma = 300$~\kms .

We generate spectra with a range of  S/N ratios (per \AA ), from 30 to
500,  by  adding  Gaussian  noise,  and  performing  100  Monte  Carlo
realizations for each S/N value. For each simulated spectrum, we apply
the  different fitting  methods  (FSF,  FIF, and  IF)  to recover  the
abundance ratios, using models with  two SSPs, where age, metallicity,
and mass  fraction of each  component are free fitting  parameters. At
each  S/N, we  quantify  the  bias and  uncertainty  in the  recovered
\xfe\ values  by computing  the median and  standard deviation  of the
differences          between         input          and         output
abundances.  Figures~\ref{fig:sims_ls} and~\ref{fig:sims_hs}  plot the
median  (dashed  lines) and  standard  deviation  (solid lines)  as  a
function    of   S/N,    for   low-    and   high-$\sigma$    spectra,
respectively. Different  fitting methods are indicated  with different
colors (as labeled  in the Figures). The results can  be summarized as
follows.
\begin{itemize}
\item -- All methods yield unbiased estimates of abundance ratios, with median residuals consistent with zero in all cases. As expected, the uncertainties increase with decreasing S/N.  
\item --  At high S/N ($>100$), all methods tend to perform similarly, with uncertainties $\lesssim 0.1$~dex. This result contrasts with CvD18, who found -- although using a different set of indices than those employed here -- that even at high S/N, IF yields much larger uncertainties than FSF.
\item --  The presence of a young component (i.e., in low-$\sigma$ ETGs) results in significantly larger uncertainties, especially for elements such as O, which are intrinsically more difficult to constrain. In contrast, high-$\sigma$ galaxies show uncertainties $\lesssim 0.1$~dex even at S/N$=30$.
\item --  FSF provides the most precise estimates (i.e., the lowest scatter), as it fully exploits the information content of the spectrum. In the absence of a young component (Figure~\ref{fig:sims_hs}), 
FIF and IF  perform similarly, while in the presence of a young component, FIF outperforms IF, as indicated by the lower standard deviations (red solid lines below blue) in Fig.~\ref{fig:sims_ls}.
\item --  For some elements, such as Mg and Ca, whose effect is strong in prominent absorption features (i.e., Mgb, Ca4227, and CaT), all methods yield  similar uncertainties as a function of S/N.
\end{itemize}
We point out that these results are based on simulations where the same SSP models are used to create the mock spectra and to fit them. In reality, due to small systematics in both the data and the models, different techniques do not necessarily yield consistent abundance ratio estimates, as shown by our analysis of the observed spectra (see Sec.~\ref{sec:results}).

\section{Quality of spectral fitting for the M31 bulge}
\label{app:fits}
\begin{figure*}
\begin{center}
 \leavevmode
 \includegraphics[width=18cm]{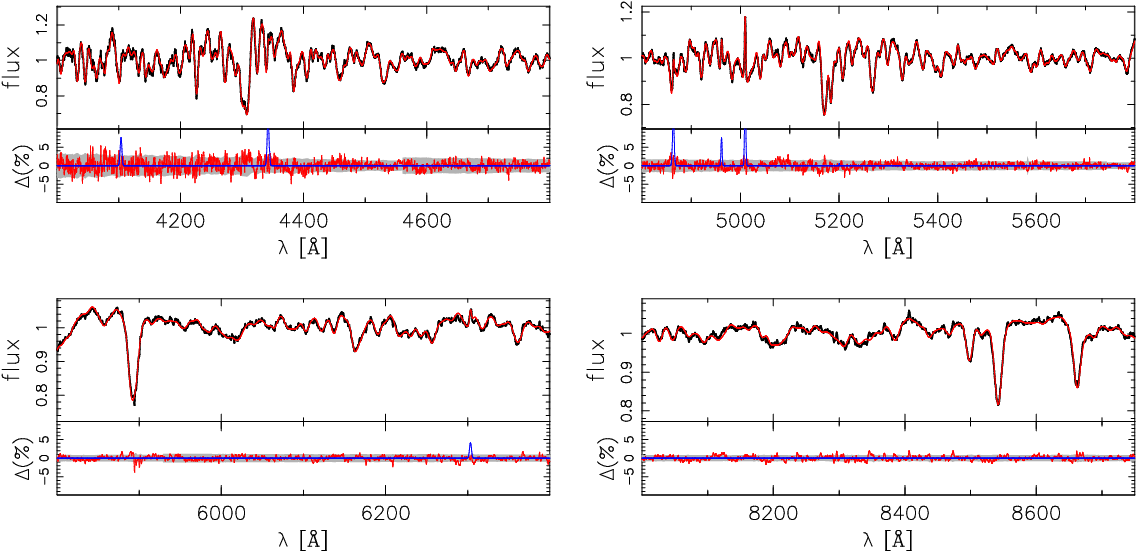}
\end{center}
 \caption{Example of full-spectral fitting for one of the M31 spectra. The four panels correspond to the spectral ranges defined in Sect.~\ref{sec:methods}. In each panel, the upper subplot shows the observed spectrum (black) and the best-fitting model (red), both normalized by the best-fitting multiplicative polynomial, while the lower subplot shows the relative residuals (observed minus model), the $\pm$1-sigma uncertainties (gray shaded regions), and the best-fitting model for the emission lines (blue curves).   
 }
   \label{fig:fsf}
\end{figure*}

Figure~\ref{fig:fsf} plots an example of FSF for one of the M31 spectra, at a radial distance of approximately $-60$\arcsec\ from the galaxy's photometric center. The fit quality is excellent, with residuals $\lesssim 1 \%$ in most pixels, and a mean absolute deviation of $\sim 0.6 \%$. Emission lines (blue curves in the lower subpanels) are modeled with comparable accuracy to the absorption features. Although the spectra have been corrected for ISM contamination in the NaD line (see~LB21), the correction is not perfect, and small residuals at the 2--3$\%$ level are visible in the trough of the NaD absorption. However, these residuals have a negligible impact on our results. Masking out the affected pixels does not significantly alter the derived \nafe\ abundance ratios, with variations $\lesssim 0.03$~dex. Fitting residuals increase slightly, by $\sim 1 \%$ on average, at wavelengths shorter than $\sim 4400$~\AA , primarily due to larger flux uncertainties in the blue part of the spectrum and, more importantly, to the higher model uncertainties in the theoretical spectral responses to varying abundance ratios { (see~\citealt{CvD12a})}.

\begin{figure*}
\begin{center}
 \leavevmode
 \includegraphics[width=18.5cm]{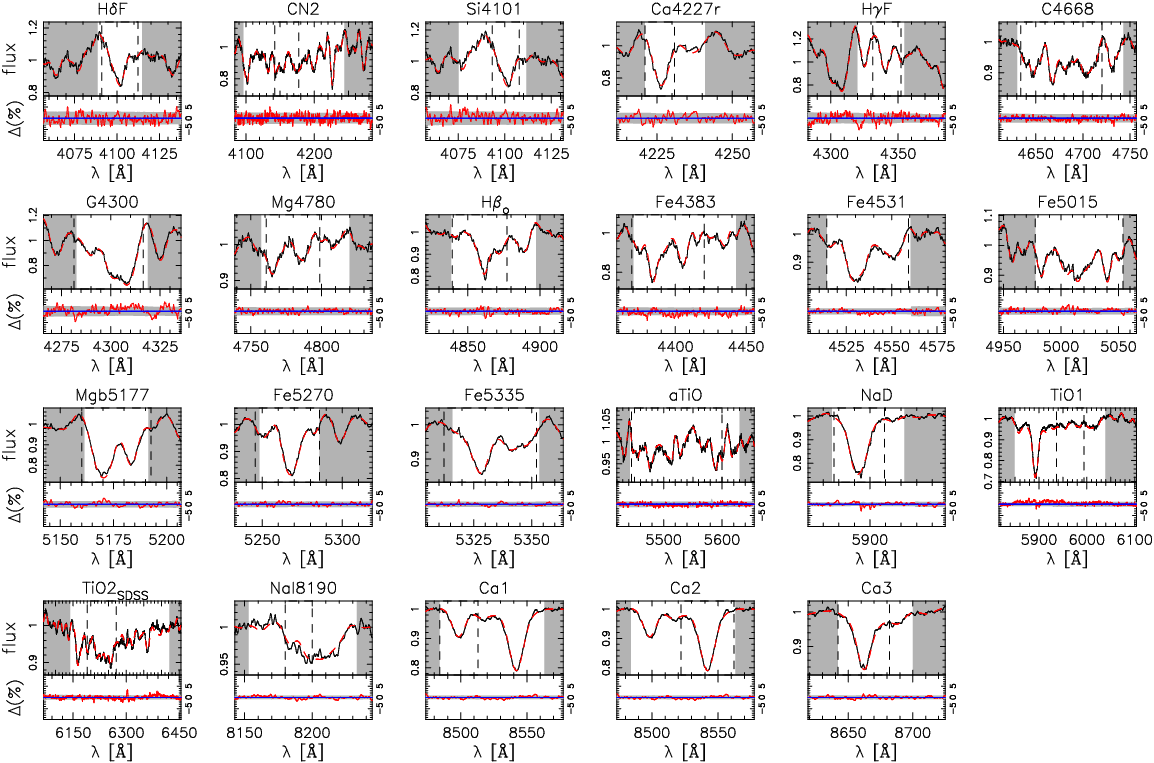}
\end{center}
 \caption{Same as Fig.~\ref{fig:fsf} but for full-index fitting. Different panels correspond to the spectral indices defined in Sect.~\ref{sec:methods}. In the upper subplots, gray shaded regions mark the index pseudocontinua, while vertical dashed lines define the index passbands.    
 }
   \label{fig:fif}
\end{figure*}

Figure~\ref{fig:fif} shows an example of FIF for the same spectrum as in Fig.~\ref{fig:fsf}. We run FIF on the spectrum cleaned for emission lines, modeled with the FSF approach (i.e., using the blue curve in the lower subpanels of Fig.~\ref{fig:fsf}). The fit quality is excellent, with residuals comparable to those obtained with FSF.

\section{Abundance ratios of ETGs from SDSS stacked spectra}
\label{app:sdss}
\begin{figure*}
\begin{center}
 \leavevmode
 \includegraphics[width=16cm]{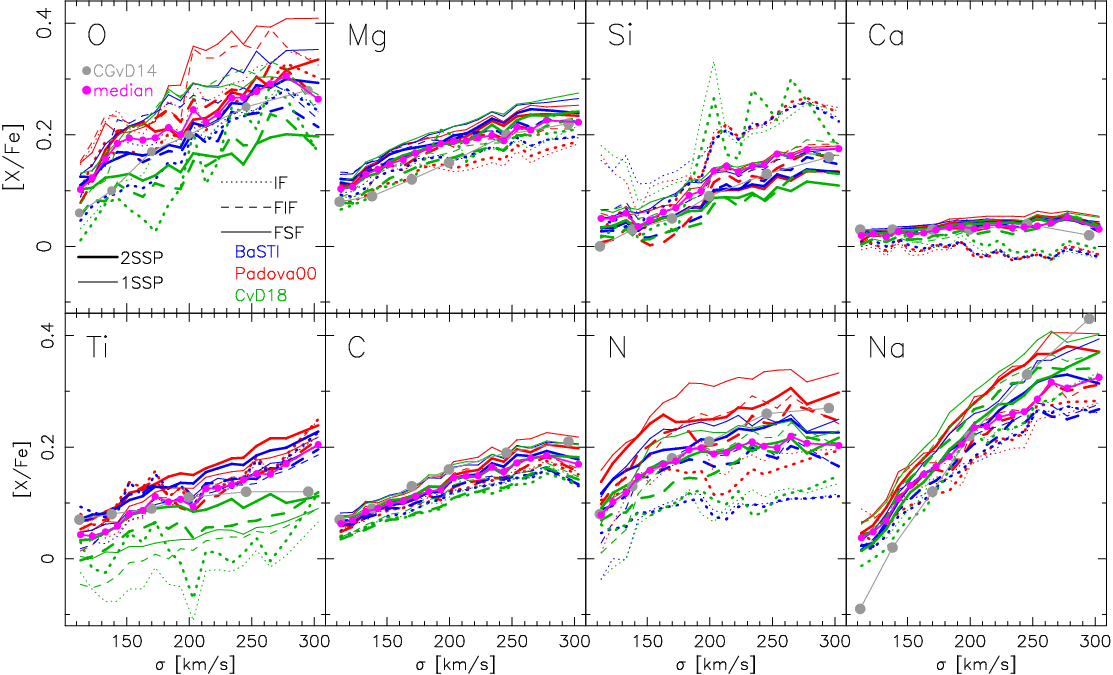}
\end{center}
 \caption{
 Abundance ratios versus velocity dispersion, $\sigma$, for SDSS stacked spectra of ETGs (see the text for details). Each panel corresponds to a different element, as labeled. Thin and thick lines show results obtained by fitting 1SSP and 2SSP models, respectively. Dotted, dashed, and solid lines denote different fitting techniques. Different models are shown in different colors, i.e., blue for E-MILES BaSTI, red for E-MILES Padova00, and green for CvD18. Gray dots show the \xfe\ versus $\sigma$ trend { from~\citet{CGvD:2014}}, based on a different set of SDSS stacked spectra. 
 }
   \label{fig:sdss_xfe}
\end{figure*}

Figure~\ref{fig:sdss_xfe} shows the trend of individual abundance ratios for SDSS ETGs as a function of velocity dispersion, $\sigma$, with different colors (line types) corresponding to different models (fitting methods). The median trends among all methods are plotted as magenta dots, and are compared with those from~\citet{CGvD:2014}, who employed a preliminary version of the CvD18 models and an FSF approach based on 2SSP models, but using an independent set of SDSS stacked spectra (gray dots). Results can be summarized as follows:
\begin{description}
\item --  Abundance ratios increase with $\sigma$, except for Ca, which closely tracks Fe at both low and high $\sigma$. A flattening in the \nfe\ trend is observed for $\sigma \gtrsim 200$~\kms . Overall, these trends are qualitatively consistent with previous studies ({ e.g.,~\citealt{CGvD:2014, JTM:2012, WTS:14, Parikh:2019, Li:25}}). 
\item --  The trends obtained using the FSF method with CvD18 and 2SSP models (see thick green solid line in the Figure) can be directly { compared} to those from CvDG14. While there is a general agreement, some interesting differences are found. Our trends for Na and O -- and, to a lesser extent, for N and Si -- are steeper than those { of~\citet{CGvD:2014}}, whereas consistent results (within $\lesssim 0.05$~dex) are found for the other elements. 
These discrepancies likely stem from differences in sample selection and the stacking procedure of SDSS spectra, as well as from the { fact { that} \citet{CGvD:2014}} used a preliminary version of CvD18 models, which did not include, for example, the effect of varying metallicity.
\item --  Small variations are observed among different methods for some elements, such as Mg, C, and Ca, though in the latter case the IF method tends to give lower abundance ratios ({ by $\lesssim 0.05$~dex).  Variations among methods are larger for other elements: for O, CvD18 (Padova00 1SSP) models give systematically lower (higher) abundances, with absolute deviations up to $\sim$0.1~dex; \sife\ is systematically higher (up $\sim$0.1~dex) for the 1SSP fits; for N, 1SSP fits (Padova00 1SSP models) give \xfe\ values deviating by $\sim 0.1$~dex from the average trend.}
For Ti, CvD18 models give systematically lower values than E-MILES, as seen from the fact that the corresponding green lines in Figure~\ref{fig:sdss_xfe} lie systematically below the other curves. This is relevant, as Ti is expected to be produced only in Type II SNe~\citep{Nomoto:2013}, and  therefore  should follow the trends of the light $\alpha$-elements (O and Mg), consistent with results from E-MILES models. Alternatively, a genuinely lower abundance of Ti in the most massive galaxies would require a modification of nucleosynthetic yields (see the discussion { in~\citealt{CGvD:2014}}). 
\end{description}

\section{Total metallicity estimates}
\label{app:mh}
For each spectrum, the fitting methods described in Sec.~\ref{sec:methods} provide the abundance variations, $\rm \delta [X/H]$, with respect to the abundance pattern of the empirical SSPs (either CvD18 or E-MILES). As these empirical models are approximately scaled-solar, in the absence of any theoretical response applied to Fe, their total metallicity \zh\ coincides with \feh , and thus the best-fitting values of $\rm \delta [X/H]$ directly correspond to the abundance ratios \xfe . However, to increase the flexibility of the fitting, we also allowed Fe to vary. This choice has only a minor impact (at the few percent level) on the inferred \xfe\ estimates. We therefore compute iron metallicity as \feh$=\rm [Z/H] + \delta [Fe/H]$, and the abundance ratios as \xfe$=\rm \delta [X/H] - \delta [Fe/H]$.

The total metallicity is then estimated using the equation:
\begin{equation}
\rm [M/H] = [Fe/H] + \log_{10} \left( \sum_X Z_X*(10^{[X/Fe]}-1)+1 \right),
\label{eq:mh}
\end{equation}  
where the sum includes all fit abundance ratios $\rm [X/Fe]$, and $\rm Z_X$ denotes the mass fraction of element $\rm X$ in a solar-composition mixture. Although total metallicity is computed through Eq.~\ref{eq:mh} in the present work, for small deviations from solar scale, Eq.~\ref{eq:mh} can also be approximated by a linear expression:
\begin{equation}
\rm [M/H] \simeq [Fe/H] + \sum_X Z_X \cdot [X/Fe].
\label{eq:mh2}
\end{equation}  
For \xfe $\lesssim 0.2$~dex, this approximation holds with an accuracy better than $\sim 0.03$~dex in \mh . Adopting the solar-composition mixture of~\citealt{Asplund:2009}, only a few elements have significant mass fractions ($\rm Z_X >1$\%), namely O, Mg, Si, C, and N, whose $\rm Z_X $ values are 0.55, 0.05, 0.05, 0.18, and 0.05, respectively. As expected, the larger contribution is that of O, while the combined mass fraction of $\alpha$-elements (O, Mg, Si, Ca, TI) amounts to $\rm \sum_\alpha Z_\alpha \simeq 0.65$. 
It is important to note that Eq.~\ref{eq:mh} provides only an approximate estimate of total metallicity, as the theoretical response functions in the CvD18 models are computed assuming solar-scaled isochrones. 
When varying all the $\alpha$ elements in lockstep, for $\rm [\alpha/Fe] \le 0.4$~dex, Eq.~\ref{eq:mh} can be approximated with an accuracy better than $0.01$~dex by the linear relation:
\begin{equation}
\rm [M/H] \simeq [Fe/H] + A \cdot [ \alpha / Fe],
\label{eq:mh3}
\end{equation}  
where the constant $\rm A$ is $\sim 0.76$ ($\sim 0.73$) for the ~\citealt{Asplund:2009} (\citealt{Grevesse:1998}) solar composition mixture, consistent with equation~4 of~\citet{Vazdekis:15}.

\section{Age estimates}
\label{app:ages}
\begin{figure}
\begin{center}
 \leavevmode
 \includegraphics[width=8.2cm]{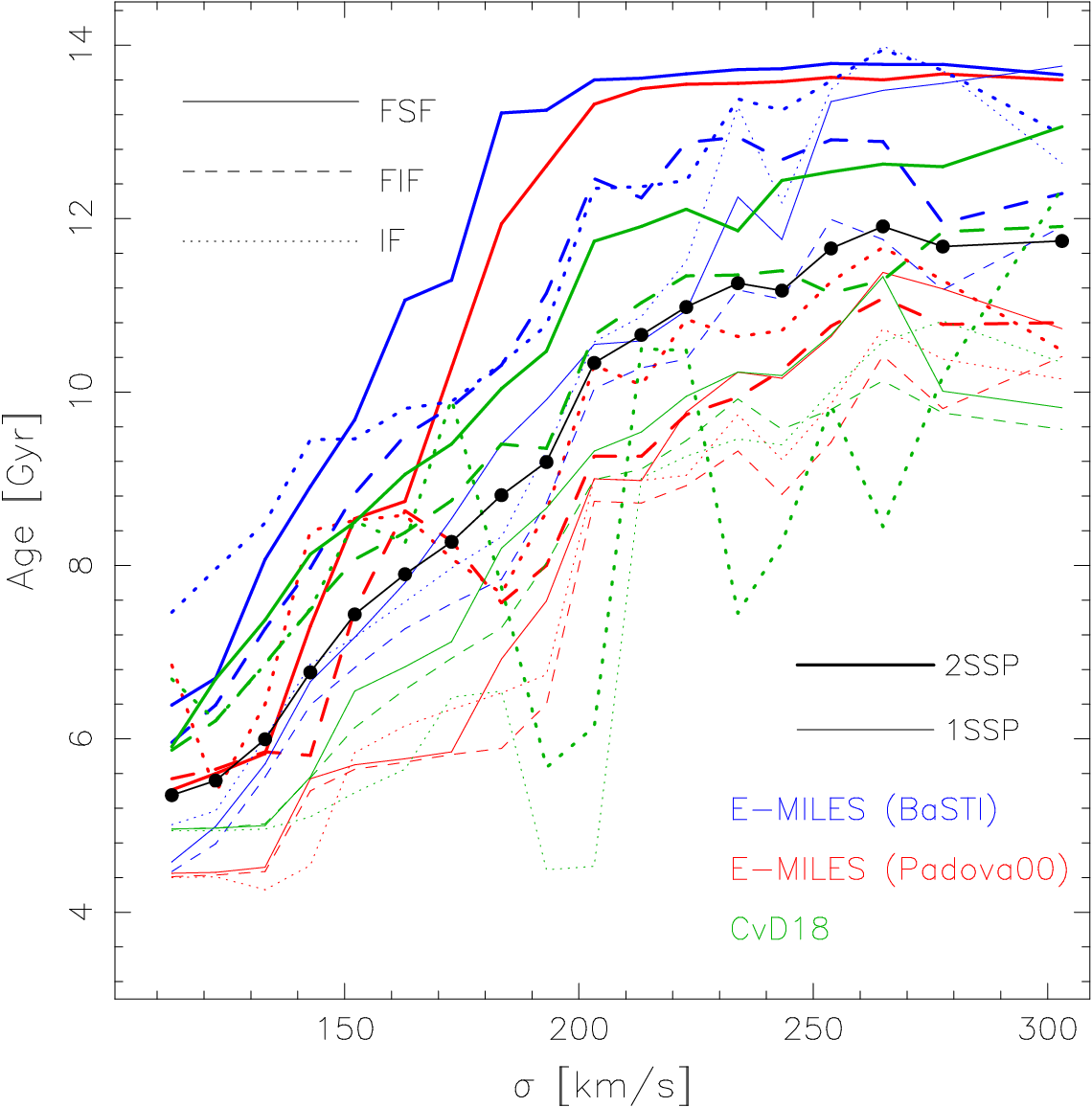}
\end{center}
 \caption{
 Same as Fig.~\ref{fig:xh_sdss} but plotting r-band luminosity-weighted age of ETGs stacked spectra as a function of velocity dispersion, $\sigma$.
 }
   \label{fig:ages_sdss}
\end{figure}

\begin{figure}
\begin{center}
 \leavevmode
 \includegraphics[width=8.2cm]{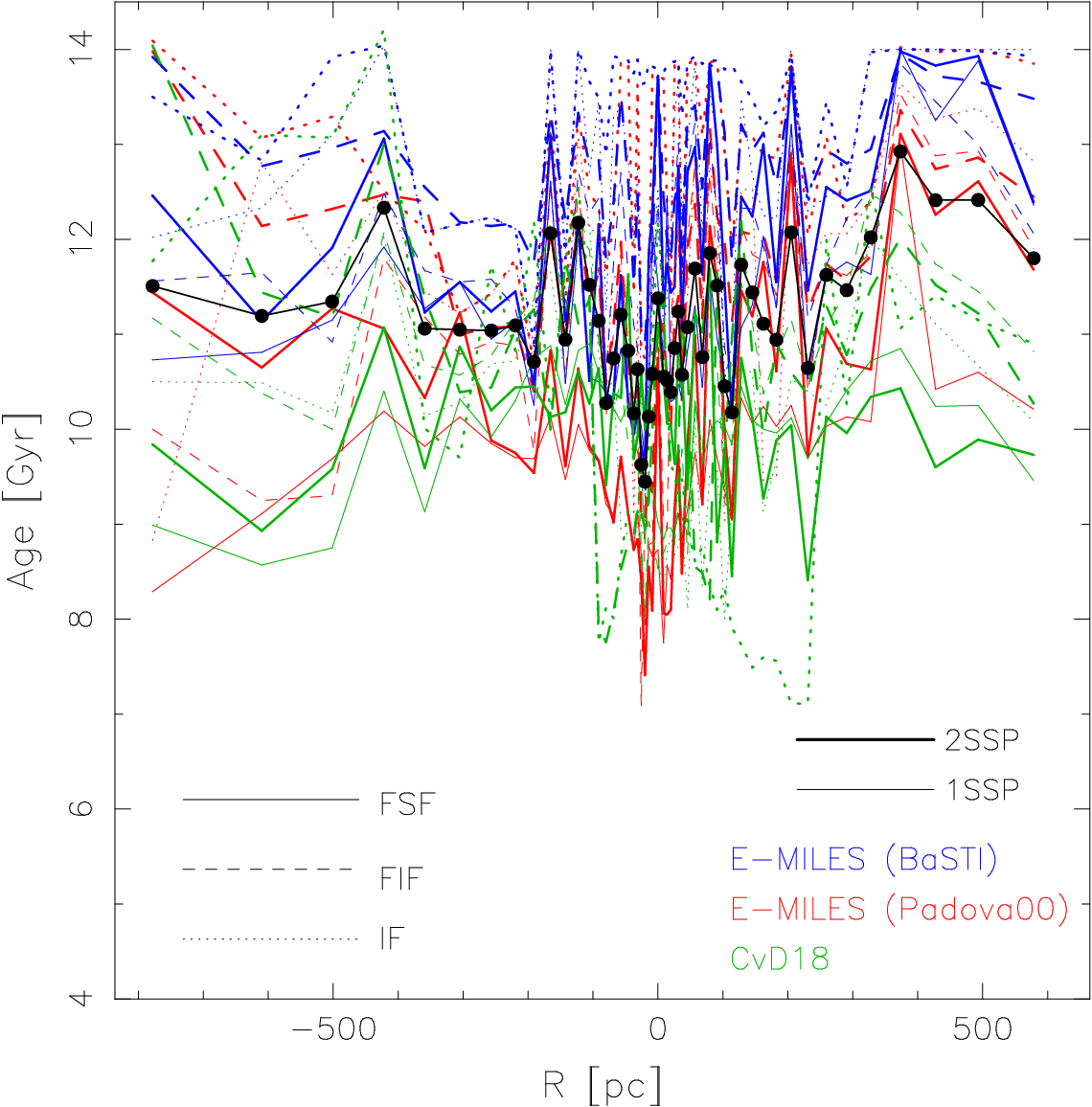}
\end{center}
 \caption{
 Same as Fig.~\ref{fig:xh_m31} but plotting r-band luminosity-weighted age as a function of galactocentric distance, R, for the bulge of M31. 
 }
   \label{fig:ages_m31}
\end{figure}

{ 

Fig.~\ref{fig:ages_sdss} plots the r-band luminosity-weighted age as a function of velocity dispersion for the SDSS stacked spectra of ETGs from different fitting methods. Consistent with LB13 and previous studies ({ e.g.,~\citealt{CGvD:2014, Thomas:2010}}), we find that the age increases with $\sigma$, with a large dispersion among different methods at high $\sigma$. 

For the bulge of M31 (Fig.~\ref{fig:ages_m31}), we find predominantly old ages (on average $\sim$11.5~Gyr) with a significant scatter (about $\pm 2$~Gyr across the different methods) and evidence for a mildly younger central population (by $\sim$2~Gyr). These results agree with previous studies (see~LB21 and references therein). Remarkably, the bulge age resembles that of massive ETGs, reinforcing our conclusion that its formation was similar to that of the most massive galaxies.

We note that, to reproduce the bulge abundance ratios, our chemical-evolution models require a very short star-formation timescale (Sec.~\ref{sec:models}). Consequently, the models do not provide robust age predictions across different regions of the bulge and therefore cannot be directly compared with the observations.

}

\section{Predictions of chemical evolution models}
\label{app:MDF}
\begin{figure}
\begin{center}
 \leavevmode
 \includegraphics[width=8.2cm]{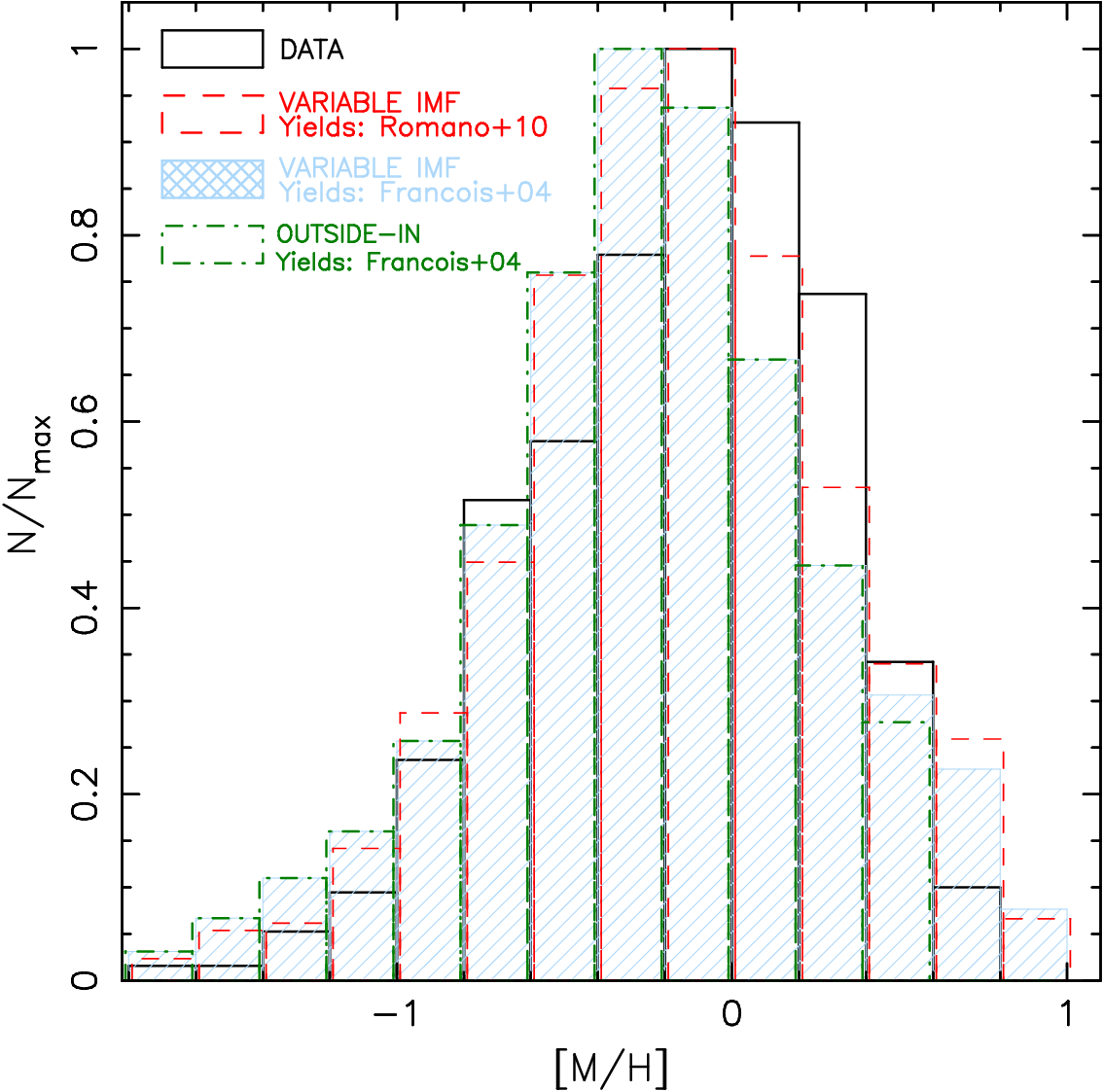}
\end{center}
 \caption{Comparison of the observed metallicity distribution function of the M31 bulge from~\citet{SJ:2005} (black) with that predicted by our chemical evolution models for a variable IMF scenario, adopting yields from R10  and F04, shown as dashed red and hatched light-blue histograms, respectively, and an outside-in model with yields from F04 (dot-dashed green histogram).
 Slight shifts have been applied to the histograms for display purposes.
 }
   \label{fig:MDF}
\end{figure}

Figure~\ref{fig:MDF} compares the observed MDF from SJ05 with that predicted by our chemical evolution models, which assume an intense and rapid formation of the bulge. We consider models for variable IMF scenario, adopting yields from R10  and F04, as well as an outside-in model with yields from F04.
We note that the MDF of SJ05 was derived by fitting the color-magnitude diagram (CMD) of RGB stars in the bulge at a distance of $\sim$1.6~kpc from the galaxy center, whereas the largest radius probed by our models is $\sim$600~pc. This difference in galactocentric distance should not significantly affect our conclusions, as both the metallicity and abundance ratio profiles in the bulge flatten at distances greater than $\sim$100~pc from the center. In general, we find reasonably good agreement between the observed and model MDFs, supporting the physical parameters adopted in our chemical evolution models (Sec.~\ref{sec:cmodels}).

\begin{figure}
\begin{center}
 \leavevmode
 \includegraphics[width=8.2cm]{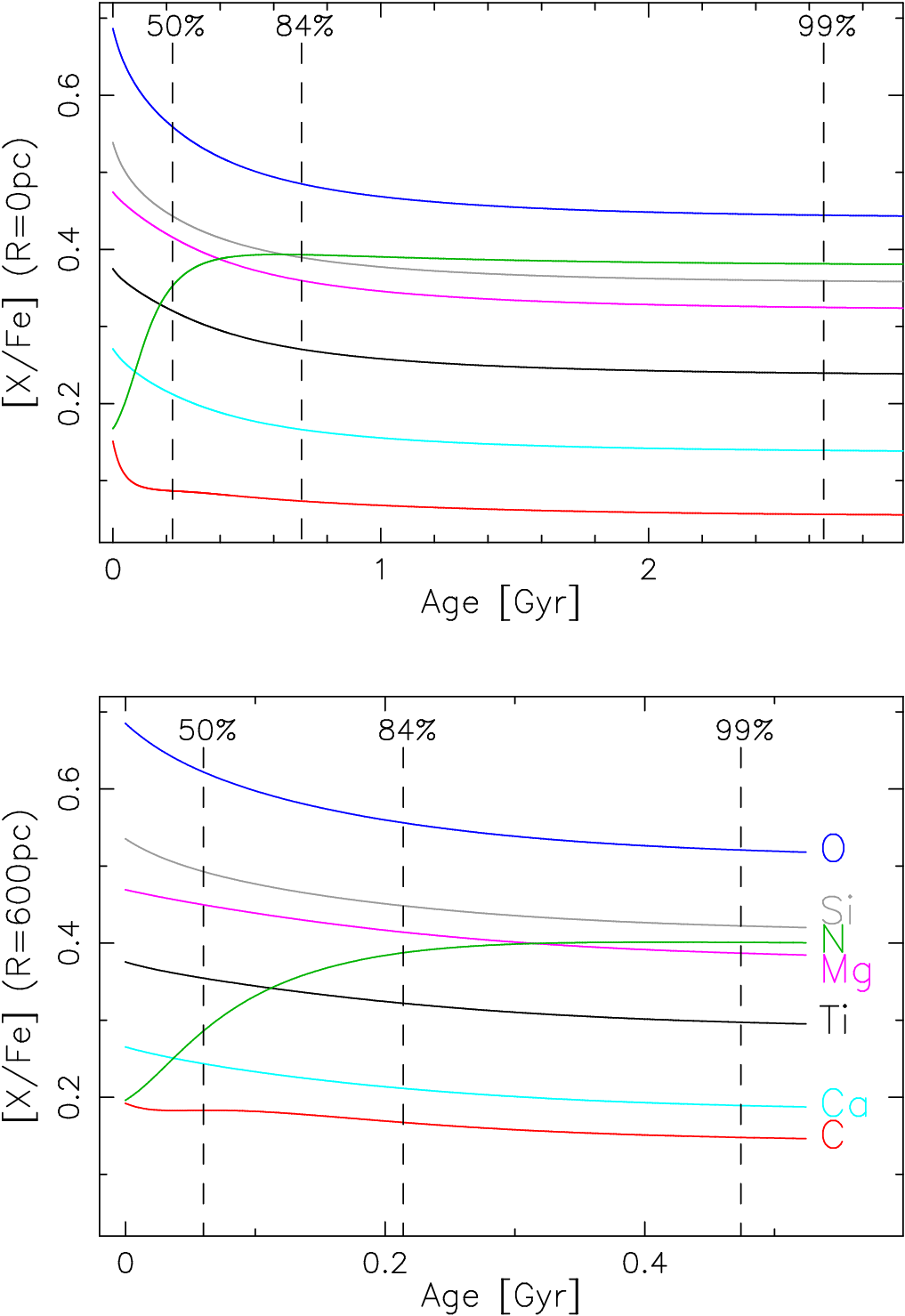}
\end{center}
 \caption{
 Examples of the evolution of abundance ratios with age for our outside-in chemical evolution model based on F04 yields.
 The top and bottom panels show results at $\rm R=0$ and $\rm R=600$~pc, respectively (see Tab.~\ref{tab_models}). Different colors correspond to different elements (see bottom panel), as in Fig.~\ref{fig:xfe_chem}. Dashed vertical lines mark the ages where the models reach 50, 84, and 99$\%$ of the total stellar mass formed. In the bottom panel, the curves end at the age where the cumulative mass fraction reaches 100$\%$.
 }
   \label{fig:xfe_age}
\end{figure}

{ 
Figure~\ref{fig:xfe_age} presents the evolution of abundance ratios as a function of age in our outside-in chemical evolution model based on the F04 yields, for the innermost and outermost galactocentric distances considered ($R=0$ and $R=600$~pc, respectively). 
Dashed vertical lines mark the ages at which 50, 84, and 99$\%$ of the total stellar mass has formed. Sodium abundances are not included in the figure as the corresponding yields are only available for the R10 model. In the central region, where star formation is more extended, abundance ratios remain nearly constant after $\sim$1~Gyr, whereas in the outermost region they reach a plateau much earlier, after $\sim$0.3~Gyr. This behavior reflects the more rapid chemical evolution at larger galactocentric distances.
}

\end{appendix}

\end{document}